\documentclass[letterpaper,UKenglish]{lipics}
 
\title{Edge-Unfolding Nearly Flat Convex Caps}

\author[1]{Joseph O'Rourke}
\affil[1]{Department of Computer Science, Smith College, Northampton, MA, USA\\
  \texttt{jorourke@smith.edu}}

\authorrunning{J.\,O'Rourke} 

\Copyright{Joseph O'Rourke}

\subjclass{F.2.2 Nonnumerical Algorithms and Problems. G.2.2 Graph Theory. }
\keywords{polyhedra, unfolding}

\usepackage{tcolorbox} 
\theoremstyle{plain}
\newtheorem{proposition}[theorem]{Proposition}
\newcommand{\lemlab}[1]{\label{lemma:#1}}
\newcommand{\thmlab}[1]{\label{thm:#1}}
\newcommand{\proplab}[1]{\label{prop:#1}}
\newcommand{\corlab}[1]{\label{cor:#1}}
\newcommand{\figlab}[1]{\label{fig:#1}}
\newcommand{\seclab}[1]{\label{sec:#1}}

\newcommand{\lemref}[1]{\ref{lemma:#1}}
\newcommand{\thmref}[1]{\ref{thm:#1}}
\newcommand{\corref}[1]{\ref{cor:#1}}
\newcommand{\propref}[1]{\ref{prop:#1}}
\newcommand{\secref}[1]{\ref{sec:#1}}
\newcommand{\figref}[1]{\ref{fig:#1}}

\def\P{{\mathcal P}}
\def\Q{{\mathcal Q}}
\def\C{{\mathcal C}}
\def\bcC{{\partial \mathcal C}}
\def\bC{{\partial C}}
\def\E{{\mathcal E}}
\def\S{{\mathcal S}}
\def\T{{\mathcal T}}

\def\cF{{\mathcal F}}
\def\g{{\gamma}}
\def\f{{\phi}}
\def\F{{\Phi}}
\def\l{{\lambda}}
\def\r{{\rho}}
\def\L{{\Lambda}}
\def\o{{\omega}}
\def\O{{\Omega}}
\def\D{{\Delta}}

\def\e{{\varepsilon}}
\def\a{{\alpha}}
\def\b{{\beta}}
\def\q{{\theta}}

\def\t{{\tau}}
\def\s{{\sigma}}
\def\R{{\mathbb{R}}}

\usepackage{color}

\newcommand{\remove}[1]{{}}

\begin{document}

\maketitle

\begin{abstract}
The main result of this paper is a proof that a
nearly flat, acutely triangulated convex cap $\C$ in $\R^3$
has an edge-unfolding to a non-overlapping polygon in the plane.
A \emph{convex cap} is the intersection of the surface of a convex polyhedron and a halfspace.
``Nearly flat'' means that every outer face normal forms a 
sufficiently small angle $\f < \F$ with the $\hat{z}$-axis orthogonal to the halfspace bounding plane.
The size of $\F$ depends on the acuteness gap $\a$:
if every triangle angle is at most $\pi/2 {-} \a$,
then $\F \approx 0.36 \sqrt{\a}$ suffices; e.g., for $\a=3^\circ$,
$\F \approx 5^\circ$.
Even if $\C$ is closed to a polyhedron by adding
the convex polygonal base under $\C$,
this polyhedron can be edge-unfolded without overlap.
The proof employs the recent concepts of
angle-monotone and radially monotone curves.
The proof is constructive, leading to a polynomial-time algorithm
for finding the edge-cuts, at worst $O(n^2)$; a version has been implemented.
\end{abstract}

\section{Introduction}
\seclab{Introduction}
Let $\P$ be a convex polyhedron in $\R^3$, and let $\phi(f)$  be
the angle the outer normal to face $f$ makes with the $\hat{z}$-axis.
Let $H$ be a halfspace whose bounding plane is orthogonal to the $\hat{z}$-axis, and includes points
vertically above that plane.
Define a \emph{convex cap} $\C$ of angle $\F$ to be $C=\P \cap H$
for some $\P$ and $H$, such that $\f(f) \le \F$ for all $f$ in $\C$.
We will only consider $\F < 90^\circ$, which implies that the projection
$C$ of $\C$ onto the $xy$-plane is one-to-one.
Note that $\C$ is not a closed polyhedron; it has no ``bottom,''
but rather a boundary $\bcC$.

Say that a convex cap $\C$ is \emph{acutely triangulated}
if every angle of every face is strictly acute, i.e., less than $90^\circ$.
Note that $\P$ being acutely triangulated does not imply that 
$\C=\P \cap H$ is acutely triangulated.
It may be best to imagine first constructing $\P \cap H$ and then 
acutely triangulating the surface.
That every polyhedron may be acutely triangulated was first established by
Burago and Zalgaller~\cite{bz-pen-60}. 
Recently Bishop proved that every PSLG (planar straight-line graph) 
of $n$ vertices
has a conforming acute triangulation, using $O(n^{2.5})$ triangles~\cite{bishop2016non-obtuse}.%
\footnote{
His main Theorem~1.1 is stated for non-obtuse triangulations, but he says later that ``the theorem
also holds with an acute triangulation, at the cost of a larger constant in the $O(n^{2.5})$.''}
Applying Bishop's algorithm will create edges with
flat ($\pi$) dihedral angles, resulting
from partitioning an obtuse triangle into several acute triangles.
One might view the acuteness assumption as adding extra possible cut edges.

An \emph{edge-unfolding} of a convex cap $\C$ is a cutting of edges of
$\C$ that permits $\C$ to be developed to the plane as a simple
(non-self-intersecting) polygon, a ``net.''
The cut edges must form a boundary-rooted spanning forest $\cF$:
a forest of trees, each rooted on the boundary rim $\bcC$, and spanning
the internal vertices of $\C$.
Our main result is: 
\begin{theorem}
Every acutely triangulated convex cap $\C$
with face normals bounded by a sufficiently small angle $\F$
from the vertical,
has an edge-unfolding to a non-overlapping polygon in the plane.
The angle $\F$ is a function of the acuteness gap $\a$
(Eq.~\ref{eq:Dq}).
The cut forest can be found in quadratic time.
\thmlab{CCapEUnf}
\end{theorem}
\noindent
\noindent
An example is shown in 
Fig.~\figref{AM_25_35_s2_v98_Lay3D},
and another example in the Appendix, Fig.~\figref{AM_20_30_s2_v83_ell_Lay3D}.
\begin{figure}[htbp]
\centering
\includegraphics[width=1.0\linewidth]{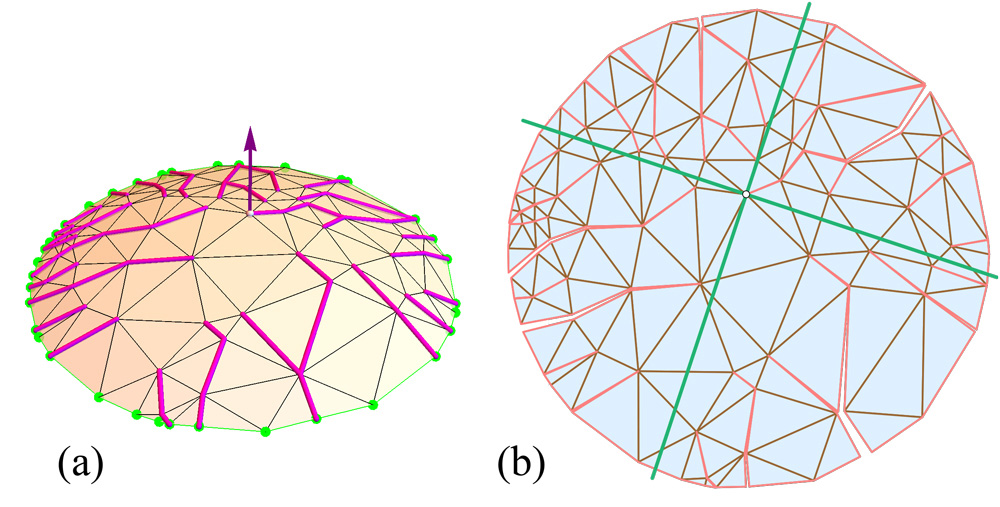}
\caption{(a)~A convex cap of $98$ vertices, $\F \approx 33^\circ$, with spanning forest $\cF$ marked. 
$\C$ is non-obtusely triangulated (rather than acutely triangulated).
(b)~Edge-unfolding by cutting $\cF$.
The quadrant lines are explained in Section~\protect\secref{SpanningForest}.}
\figlab{AM_25_35_s2_v98_Lay3D}
\end{figure}

\subsection{Background}
It is a long standing open problem whether or not every convex polyhedron
has a non-overlapping edge-unfolding, often called 
D\"urer's problem~\cite{do-gfalop-07}~\cite{o-dp-2013}.
Theorem~\thmref{CCapEUnf} can be viewed as an advance on a narrow
version of this problem.
This theorem---without the acuteness assumption---has been a folk-conjecture 
for many years.
A specific line of attack
was conjectured in~\cite{lo-ampnt-17}, which obtained a result
for planar non-obtuse triangulations,
and it is that sketch I follow for the proof here.

There have been two recent advances on D\"urer's problem.
The first is Ghomi's positive result that sufficiently thin polyhedra have edge-unfoldings~\cite{g-aucp-14}. This can be viewed as a counterpart to 
Theorem~\thmref{CCapEUnf}, which when supplemented by~\cite{o-aeucc-17}
shows that sufficiently flat polyhedra 
have edge-unfoldings.
The second is a negative result that shows that when restricting cutting to geodesic
``pseudo-edges'' rather than edges of the polyhedral skeleton,
there are examples that cannot avoid overlap~\cite{barvinok2017pseudo}.

It is natural to hope that Theorem~\thmref{CCapEUnf} might lead to an edge-unfolding result
for all acutely triangulated convex polyhedra,
but I have been so far unsuccessful in realizing this hope.
Possible extensions are discussed in Section~\secref{Discussion}.

\section{Overview of Algorithm}
\seclab{OverviewAlgorithm}
We now sketch the simple algorithm in four steps; the proof of correctness
will occupy the remainder of the paper.
First, $\C$ is projected orthogonally to $C$ in the $xy$-plane,
with $\F$ small enough so that the acuteness gap of $\a > 0$ decreases
to $\a' \le \a$ but still $\a' > 0$. So $C$ is acutely triangulated.
Second, a boundary-rooted angle-monotone spanning forest $F$
for $C$ is found using the algorithm in~\cite{lo-ampnt-17}.
Both the definition of angle-monotone and the algorithm will be
described in Section~\secref{AngMonoForest} below, but for now
we just note that each leaf-to-root path in $F$ is both $x$- and $y$-monotone
in a suitably rotated coordinate system.
Third, $F$ is lifted to a spanning forest $\cF$ of $\C$, and the
edges of $\cF$ are cut.
Finally, the cut $\C$ is developed flat in the plane.
In summary: project, lift, develop.

I have not pushed on algorithmic time complexity, but certainly $O(n^2)$ suffices,
as detailed in the full version~\cite{o-eunfcc-17}.

\section{Overview of Proof}
\seclab{Overview}
The proof relies on two results from earlier work:
the angle-monotone spanning forest result in~\cite{lo-ampnt-17},
and a radially monotone unfolding result in~\cite{o-ucprm-16}.
However, the former result needs generalization, and the latter
is unpublished and both more and less than needed here. So those results
are incorporated and explained as needed to allow this paper to stand alone.
It is the use of angle-monotone and radially monotone curves 
and their properties that
constitute the main novelties.
The proof outline has these seven high-level steps,
expanding upon the algorithm steps:
\begin{enumerate}
\item
Project $\C$ to the plane containing its boundary rim, resulting in a
triangulated convex region $C$.
For sufficiently small $\F$, $C$ is again acutely triangulated.
\item
Generalizing the result in~\cite{lo-ampnt-17},
there is a $\q$-angle-monotone, boundary-rooted spanning forest $F$ of $C$, for $\q < 90^\circ$.
$F$ lifts to a spanning forest $\cF$ of the convex cap $\C$.
\item
For sufficiently small $\F$, both sides $L$ and $R$ 
of each cut-path $\Q$ of $\cF$ are
$\q$-angle-monotone when developed in the plane, for some $\q < 90^\circ$.
\item
Any planar angle-monotone path
for an angle $\le 90^\circ$, is radially monotone,
a concept from~\cite{o-ucprm-16}.
\item
Radial monotonicity of $L$ and $R$, and sufficiently small $\F$,
imply that $L$ and $R$ do not cross in their planar development.
This is a simplified version of a result from~\cite{o-ucprm-16},
and here extended to trees.
\item
Extending the cap $\C$ to an unbounded polyhedron $\C^\infty$
ensures that the non-crossing of each $L$ and $R$
extends arbitrarily far in the planar development.
\item 
The development of $\C$ can be partitioned into
$\q$-monotone ``strips,'' whose side-to-side development layout guarantees
non-overlap in the plane.
\end{enumerate}

Through sometimes laborious arguments, I have tried to 
quantify steps even if they are in some sense obvious.
Various quantities go to zero as $\F \to 0$. 
Quantifying by explicit calculation the dependence on $\F$
lengthens the proof considerably.
Those laborious arguments and other details are relegated to the Appendix.

\subsection{Notation}
\seclab{Notation3D}
I attempt to distinguish between objects in $\mathbb{R}^3$,
and planar projected versions of those objects, either by using
calligraphy ($\C$ in $\mathbb{R}^3$ vs. $C$ in $\mathbb{R}^2$),
or primes ($\g$ in $\mathbb{R}^3$ vs. $\g'$ in $\mathbb{R}^2$),
and occasionally both ($\Q$ vs.~$Q'$).
Sometimes this seems infeasible, in which case we use different symbols
($u_i$ in $\mathbb{R}^3$ vs. $v_i$ in $\mathbb{R}^2$).
Sometimes we use $\perp$ as a subscript to indicate projections 
or developments of lifted quantities.
The plane $H$ containing $\bcC$ is assumed to be the $xy$-plane, $z=0$.

\section{Projection Angle Distortion}
\seclab{AngDistortion}
\begin{tcolorbox}
1.~Project $\C$ to the plane containing its boundary rim, resulting in a
triangulated convex region $C$.
For sufficiently small $\F$, $C$ is again acutely triangulated.
\end{tcolorbox}
This first claim is obvious: Since every triangle angle is strictly less
than $90^\circ$, and the distortion due to projection to a plane
goes to zero as $\C$ becomes more flat, for some
sufficiently small $\F$, the acute triangles remain acute under projection.

In order to obtain a definite dependence on $\F$,
the following exact bound is derived in Appendix~\secref{AngDistortion.Appendix}.

\begin{lemma}
The maximum absolute value of the
distortion $\D_\perp$ of any angle in $\mathbb{R}^3$
projected to the $xy$-plane,
with respect to the tilt $\f$ of the plane of that angle with respect to $z$,
is given by: 
\begin{equation}
\D_\perp(\f) \;=\; \cos^{-1} \left( \frac{ \sin^2 \f }{ \sin^2 \f - 2 } \right) - \pi/2 
\; \approx \; \f^2 / 2 - \f^4 /12 +O(\f^5) \;,
\label{eq:DeltaApprox}
\end{equation}
where the approximation holds for small $\f$.
\lemlab{Distortion}
\end{lemma}
In particular, 
$\D_\perp(\F) \to 0$ as $\F \to 0$.
For example, $\D_\perp( 10^\circ )  \approx  0.9^\circ$.

\section{Angle-Monotone Spanning Forest}
\seclab{AngMonoForest}
\begin{tcolorbox}
2.~Generalizing the result in~\cite{lo-ampnt-17},
there is a $\q$-angle-monotone, boundary rooted spanning forest $F$ of $C$, for $\q < 90^\circ$.
$F$ lifts to a spanning forest $\cF$ of the convex cap $\C$.
\end{tcolorbox}

First we define angle-monotone paths, 
which originated in~\cite{dfg-icgps-15}
and were further explored in~\cite{bbcklv-gtamg-16},
and then turn to the spanning forests we need here.

\subsection{Angle-Monotone Paths}
\seclab{AngMonoPaths}
Let $C$ be a planar, triangulated convex domain,
with $\bC$ its boundary, a convex polygon. 
Let $G$ be the (geometric) graph of all the triangulation edges in $C$ and on $\bC$.

Define the \emph{$\q$-wedge} $W(\b,v)$ to be the region of
the plane 
bounded by rays at angles $\b$ and $\b + \q$ 
emanating from $v$.
$W$ is closed along (i.e., includes) both rays, and has 
angular \emph{width} of $\q$.
A polygonal path $Q=(v_0,\ldots,v_k)$ following edges of $G$ is called \emph{$\q$-angle-monotone}
(or \emph{$\q$-monotone} for short)
if the vector of every edge $(v_i, v_{i+1})$ lies in $W(\b,v_0)$
(and therefore $Q \subseteq W(\b,v_0)$), for some $\b$.
(My notation here is slightly different from the notation in~\cite{lo-ampnt-17} and earlier papers.)
Note that if $\b \ge 0^\circ$ and $\b + \q \le 90^\circ$, then a $\q$-monotone path
is both $x$- and $y$-monotone, i.e., it meets every vertical, and every horizontal line
in a point or a segment, or not at all.

\subsection{Angle-Monotone Spanning Forest}
\seclab{SpanningForest}
It was proved
in~\cite{lo-ampnt-17}
that every non-obtuse triangulation $G$ of a convex region $C$
has a boundary-rooted spanning forest $F$ of $C$, with all paths in $F$
$90^\circ$-monotone.
We describe the proof and simple construction algorithm before detailing
the changes necessary for strictly acute triangulations.

Some internal vertex $q$ of $G$ is selected, and the plane partitioned into
four $90^\circ$-quadrants $Q_0,Q_1,Q_2,Q_3$ by orthogonal lines through $q$.
Each quadrant is closed along one axis and open on its counterclockwise axis;
$q$ is considered in $Q_0$ and not in the others, so the quadrants partition the plane.
It will simplify matters if we orient the axes so that no vertex except
for $q$ lies on the axes, which is clearly always possible.
Then paths are grown within each quadrant independently, as follows.
A path is grown from any vertex $v \in Q_i$ not yet included in the forest $F_i$,
stopping when it reaches either a vertex already in $F_i$, or $\bC$.
These paths never leave $Q_i$, and result in a forest $F_i$ spanning the vertices in $Q_i$ .
No cycle can occur because a path is grown from $v$ only when $v$ is not already
in $F_i$; so $v$ becomes a leaf of a tree in $F_i$.
Then $F = F_1 \cup F_2 \cup F_3 \cup F_4$.

Because our acute triangulation is of course a non-obtuse triangulation, following
the algorithm from~\cite{lo-ampnt-17} would lead to non-obtuse angle-monotone paths,
but not the $\q$-monotone paths for $\q = 90^\circ {-} \a' < 90^\circ$
we need here.
Following that construction for $\q < 90^\circ$ fails to cover the plane,
because the ``quadrants'' leave a thin $4 \a$ angular gap.
Call the cone of this aperture $g$.
We proceed as follows.

Identify an internal vertex $q$ of $G$ so that it is possible to
orient the cone-gap $g$, apexed at $q$, so that $g$ contains no internal vertices of $G$.
See Fig.~\figref{QuadGap} for an example. Then we proceed just
as in~\cite{lo-ampnt-17}: paths are grown within each $Q_i$, forming four
forests $F_i$, each composed of $\q$-monotone paths.
\begin{figure}[htbp]
\centering
\includegraphics[width=0.4\linewidth]{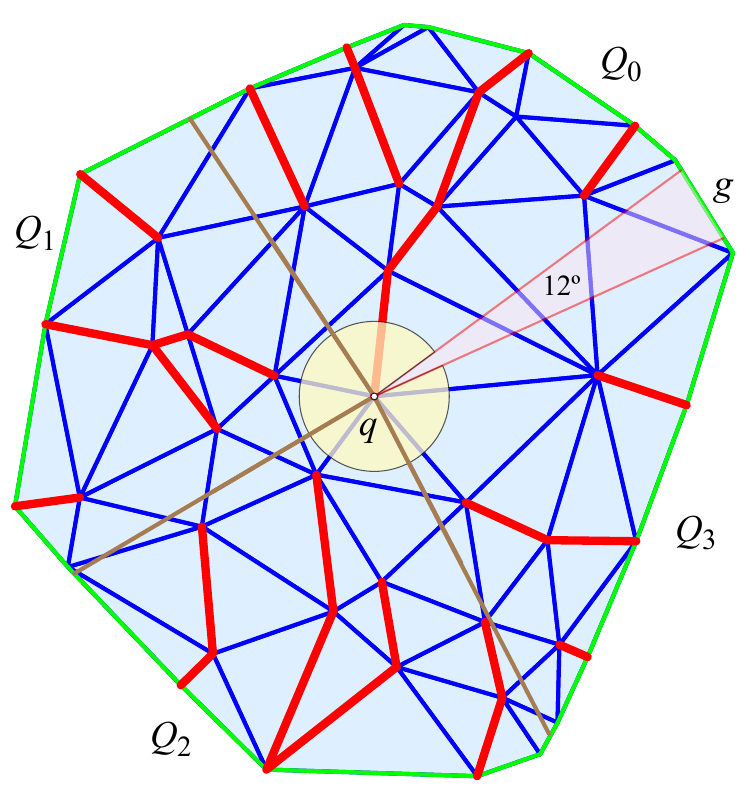}
\caption{Here the near-quadrants $Q_i$ have width $\q=87^\circ$, 
so the gap $g$ has angle $4\a  = 12^\circ$.}
\figlab{QuadGap}
\end{figure}

It remains to argue that there always is such a $q$ at which to apex cone-gap $g$.
Although it is natural to imagine $q$ as centrally located (as in Fig.~\figref{QuadGap}),
it is possible that $G$ is so dense with vertices that such a central location is not possible.
However, it is clear that the vertex $q$ that is closest to $\bC$ will suffice: aim $g$
along the shortest path from $q$ to $\bC$. Then $g$ might include several vertices on $\bC$,
but it cannot contain any internal vertices of $G$, as they would be closer to $\bC$.
Again we could rotate the axes slightly so that no vertex except for $q$ lies on an axis.

We conclude this section with a lemma:
\begin{lemma}
If $G$ is an acute triangulation of a convex region $C$,
with acuteness gap $\a'$,
then there exists a 
boundary-rooted spanning forest $F$ of $C$, with all paths in $F$
$\q$-angle-monotone, for $\q = 90^\circ {-} \a < 90^\circ$.
\lemlab{QuadGap}
\end{lemma}


\section{Curve Distortion}
\seclab{CurveDistortion}
\begin{tcolorbox}
3.~For sufficiently small $\F$, both sides $L$ and $R$ 
of each cut-path $\Q$ of $\cF$ are
$\q$-angle-monotone when developed in the plane, for some $\q < 90^\circ$.
\end{tcolorbox}
This step says, essentially, that each $\q$-monotone path $Q'$ in the planar projection
is not distorted much when lifted to $\Q$ on $\C$. 
This is obviously true as $\F \to 0$, but it requires proof. We need to establish that
the left and right incident angles of the cut $\Q$ develop to the plane as still 
$\q$-monotone paths for some (larger) $\q \le 90^\circ$.

First we bound the total curvature of $\C$ to address the phrase,
``For sufficiently small $\F$, ...''
The ``near flatness'' of the convex cap $\C$ is controlled by $\F$, the maximum
angle deviation of face normals from the $\hat{z}$-axis vertical.
Let $\o_i$ be the curvature at internal vertex $u_i \in \C$
(i.e., $2\pi$ minus the sum of the incident angles to $u_i$),
and $\O = \sum_i \o_i$ the total curvature.
In this section we bound $\O$ as a function of $\F$.
The reverse is not possible: even a small $\O$ could be realized with 
large $\F$. The bound is given in the following lemma:
\begin{lemma}
The total curvature $\O = \sum_i \o_i$ of $\C$ 
satisfies
\begin{equation}
\O \le 2 \pi ( 1 - \cos \F ) \approx \pi  \F^2-\pi \F^4/12 +O(\F^5) \;. \label{eq:OmegaPhi}
\end{equation}
\lemlab{Omega}
\end{lemma}
\vspace{-12pt}
\noindent
The proof of this lemma is in Appendix~\secref{CurveDistortion}.

Our proof of limited curve lifting distortion uses the Gauss-Bonnet theorem,%
\footnote{
See, for example, Lee's description~\cite[Thm.9.3, p.164]{lee2006riemannian}.
My $\t$ is Lee's $\kappa_N$.}
in the form $\t + \o = 2 \pi$:
the turn of a closed curve plus the curvature enclosed is $2 \pi$.

To bound the curve distortion of $Q'$, we need to bound the distortion of pieces
of a closed curve that includes $Q'$ as a subpath.
Our argument here is not straightforward, but the conclusion is
that, as $\F \to 0$, the distortion also $\to 0$:

\begin{lemma}
The difference in the total turn of any prefix of $\Q$ on the surface $\C$
from its planar projection $Q'$ is
bounded by $3 (\D_\perp + 2\O)$ (Eq.~\ref{eq:TurnQ}), which, for small $\F$, is
a constant times $\F^2$ (Eq.~\ref{eq:TurnApprox}).
Therefore, this turn goes to zero as $\F \to 0$.
\lemlab{QTurn}
\end{lemma}

The reason the proof is not straightforward is that $Q'$ could have an arbitrarily
large number $n$ of vertices, so bounding the angle distortion at each
by $\D_\perp$ would lead to arbitrarily large distortion $n \D_\perp$.
The same holds for the rim. 
So global arguments that do not cumulate errors seem necessary.

First we need a simple lemma, which is essentially the triangle
inequality on the $2$-sphere, and also proved in Appendix~\secref{AngleLifting.Appendix}.
Let $R' = \bC$ and $R = \bcC$ be the rims of the planar $C$ and of 
the convex cap $\C$, respectively. 
Note that $R'=R$ geometrically, but we will focus on the neighborhoods of these
rims on $C$ and $\C$, 
which are different.
\begin{lemma}
The planar angle $\psi'$ at a vertex $v$ of the rim $R'$ lifts to 
3D angles of the triangles of the cap $\C$ incident to $v$, 
whose sum $\psi$ satisfies  $\psi \ge \psi'$.
\lemlab{AngleLifting}
\end{lemma}

Now we use Lemma~\lemref{AngleLifting} to bound 
the total turn of the rim $R$ of $\C$ and $R'$ of $C'$.
Although the rims are geometrically identical, their turns are not.
The turn at 
vertex $a'$ of the planar rim $R'$ is $\pi - \psi'$, 
while the turn at each vertex $a$ of the 3D rim $R$ is $\pi - \psi$. 
By Lemma~\lemref{AngleLifting}, $\psi \ge \psi'$, so the turn at
each vertex of the 3D rim $R$ is 
at most the turn at each
vertex of the 2D rim $R'$. Therefore the total turn of the 3D rim $\t_{R}$
is smaller than or equal to the total turn of the 2D rim $\t_{R'}$.
And Gauss-Bonnet allows us to quantify this:
$$\t_{R'} = 2 \pi \;,\; \t_{R} + \O = 2 \pi \;,\; \t_{R'}-\t_{R} = \O \;.$$
For any subportion of the rims $r' \subset R' ,\,r \subset R$, $\O$ serves as an upper bound,
because we know the sign of the difference is the same at every vertex of $r',r$:
\begin{equation}
\t_{r'}-\t_{r} \le \O. \label{eq:rims}
\end{equation}
We can make this inference from $R$ to $r \subset R$
because of our knowledge of the signs.
Were the signs unknown, cancellation would have prevented this inference.

\subsection{Turn Distortion of $Q'$}
\seclab{TurnDistortion}
We need to bound $\D Q = | \t_Q' - \t_\Q |$,
the turn difference
between $Q'$ in the plane and $\Q$ on the surface of $\C$,
for $Q'$
any prefix of an angle-monotone path in $C$ that lifts to $\Q$ on $\C$.
The reason for the prefix here is that we want to bound the turn of any segment of $Q'$,
not just the last segment, whose turn is $\sum_i \t_i$.
And note that there can be cancellations among
the $\t_i$ along $Q'$, as we have no guarantee that they are all the same sign. 
So we take a somewhat complex approach.
Appendix~\secref{TurnDistortion.Appendix} offers a ``warm-up'' for the calculation
below.


First we sketch the situation if $Q$ cut all the way across $\C$, 
as illustrated in Fig.~\figref{GammaTurn}(a).
We apply the
Gauss-Bonnet theorem: $\t + \o = 2 \pi$, where $\o \le \O$ is the total
curvature inside the path $Q \cup r$, 
and then the planar projection (Fig.~\figref{GammaTurn}(b)), we have:
\begin{eqnarray}
\t  + \o & = & \t_{Q} + ( \t_{a} + \t_{b} ) + \t_{r} + \o 
\; = \;   2 \pi  \\ \nonumber 
\t  + \o & = & \t_{Q'} + ( \t_{a'} + \t_{b'} ) + \t_{r'} + 0 
\; = \;  2 \pi  \nonumber
\end{eqnarray}
Subtracting these equations will lead to a bound on 
$\D Q$, 
as shown in Appendix~\secref{TurnDistortion.Appendix}.


\begin{figure}[htbp]
\centering
\includegraphics[width=0.70\linewidth]{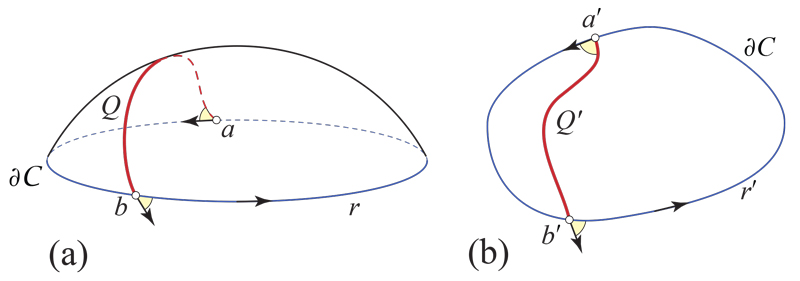}
\caption{(a)~$C$, the projection of the cap $\C$.
(b)~$\Q$ is the lift of $\Q'$ to $\C$.}
\figlab{GammaTurn}
\end{figure}

But, as indicated, $Q$ does not cut all the way across $\C$, and we need to 
bound $\D Q$ for any prefix of $Q$ (which we will still call $Q$).
Let $Q$ cut from $a \in \C$ to $b \in \bcC$.
We truncate $\C$ by intersecting with a halfspace whose bounding
plane $H$ includes $a$, as in Fig.~\figref{Trunc_n500_s1_pi3}(a).
It is easy to arrange $H$ so that $H \cap Q = \{a\}$, i.e., so that $H$ does not otherwise
cut $Q$, as follows. First, in projection, $Q'$ falls inside $\overline{W}(\q,a')$, the backward
wedge passing through $a'$. Then start with $H$ vertical and tangent to this wedge at $a$, and rotate it out to reaching $\bcC$ as illustrated. The result is a truncated cap $\C_T$.
We connect $a$ to a point $c$ on the new $\bcC_T$, depicted abstractly in 
Fig.~\figref{Trunc_n500_s1_pi3}(b).
Now we perform the analogous calculation for the curve $Q \cup r_1 \cup ca$ on $\C$,
and $\Q'  \cup r'_1 \cup ca'$:
\begin{figure}[htbp]
\centering
\includegraphics[width=0.80\linewidth]{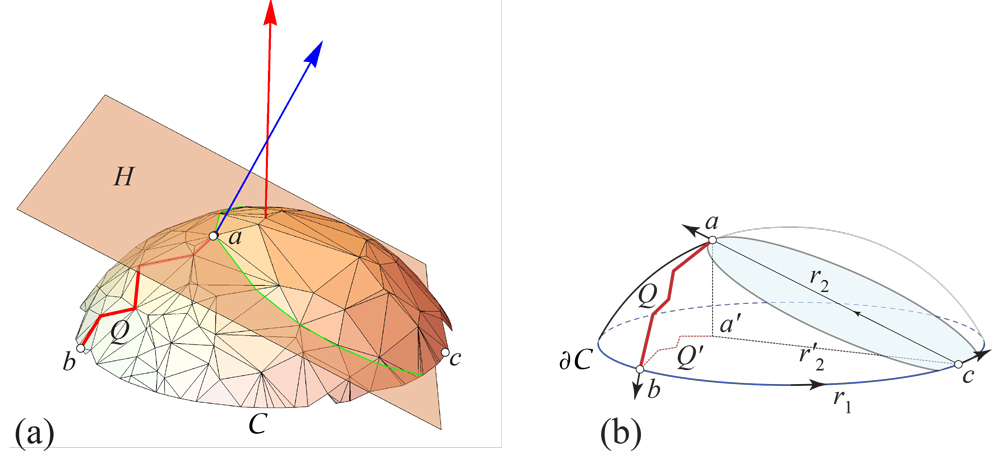}
\caption{(a)~Truncating $\C$ with $H$ so that $H \cap \Q = \{a\}$.
(b)~$r_2 = ac$ and $r'_2 = a'c$.}
\figlab{Trunc_n500_s1_pi3}
\end{figure}

\begin{eqnarray*}
\t_{Q'} + (\t_{a'} + \t_{b'} + \t_{c'}) + (\t_{r'_1}  + \t_{r'_2}) + 0 & = & 2 \pi \\ \label{eq:turnQ1}
\t_{\Q} + (\t_{a} + \t_{b} + \t_{c} )+ (\t_{r_1}  + \t_{r_2}) + \o & = & 2 \pi \label{eq:turnQ2}
\end{eqnarray*}
Subtracting leads to
\begin{eqnarray}
\t_{Q'} - \t_{\Q}  & = & 
\left( (\t_{a}{-}\t_{a'}) + (\t_{b}{-}\t_{b'}) + (\t_{c}{-}\t_{c'}) \right) + (\t_{r_1}{-}\t_{r'_1} ) + (\t_{r_2}{-}\t_{r'_2} ) + \o  \nonumber \\
\D Q & \le & 3 \D_\perp + 2 \O   \label{eq:TurnQ}
\end{eqnarray}
The logic of the bound is:
(1)~Each of the turn distortions at $a,b,c$ is at most $\D_\perp$.
(2)~The $r_1$ turn difference is bounded by $\o \le \O$.
And (3)~$\t_{r_2} = \t_{r'_2} = 0$.
Using the small-$\F$ bounds derived
earlier in Eqs.~\ref{eq:DeltaApprox} and~\ref{eq:OmegaPhi}:
\begin{equation}
|\D Q|  \; \le \; 3 \D_\perp +  2 \O
\; \approx  \; (2 \pi + \tfrac{3}{2} ) \F^2  \;. \label{eq:TurnApprox}
\end{equation}
Thus we have
$\D Q \to 0$ as $\F \to 0$, as claimed.

We finally return to the claim at the start of this section:
For sufficiently small $\F$, both sides $L$ and $R$ 
of each path $\Q$ of $\cF$ are
$\q$-angle-monotone when developed in the plane, for some $\q < 90^\circ$.

The turn at any vertex of $\Q$ is determined by the incident face angles to the left
following the orientation shown in Fig.~\figref{GammaTurn}, 
or to the right reversing that orientation (clearly the curvature enclosed by either
curve is $\le \O$).
These incident angles determine the left and right planar developments, $L$ and $R$,
of $\Q$.
Because we know that $Q'$ is
$\q$-angle-monotone for $\q < 90^\circ$,
there is some finite ``slack'' $\a = 90^\circ {-} \q$.
Because Lemma~\lemref{QTurn} established a bound for any prefix of $\Q$,
it bounds the turn distortion of each edge of $\Q$, which we can arrange
to fit inside that slack.
So the bound
provided by Lemma~\lemref{QTurn} suffices to guarantee that:
\begin{lemma}
For
sufficiently small $\F$, both $L$ and $R$ remain 
$\q$-angle-monotone for some (larger) $\q$, but still $\q \le 90^\circ$.
\lemlab{DevelAngMono}
\end{lemma}

\noindent
To ensure $\q \le 90^\circ$, we need that the maximum distortion
fits into the acuteness gap:  $| \D Q | \le \a$.
Using Eq.~\ref{eq:TurnApprox} 
leads to:
\begin{equation}
\F \; \le \; \sqrt{\frac{2}{4 \pi + 3 }}\sqrt{\a} \; \approx \; 0.36 \sqrt{\a} \;. \label{eq:Dq}
\end{equation}
For example, if all triangles are acute by $\a=4^\circ$, 
then $\F \approx 5.4^\circ$ suffices.

That $F$ lifts to a spanning forest $\cF$ of the convex cap $\C$ is immediate.
What is not straightforward is establishing the requisite properties of $\cF$.

\section{Radially Monotone Paths}
\seclab{RadialMono}
\begin{tcolorbox}
4.~Any planar angle-monotone path
for an angle $\le 90^\circ$, is radially monotone,
a concept from~\cite{o-ucprm-16}.
\end{tcolorbox}
To establish this claim, and remain independent of~\cite{o-ucprm-16},
we repeat definitions in that report. 
Let $C$ be a planar, triangulated convex domain,
with $\bC$ its boundary, a convex polygon. 
Let 
$Q=(v_0,v_1,\ldots,v_k)$ be a 
simple (non-self-intersecting) directed path of edges of $C$
connecting an interior vertex $v_0$ to a boundary vertex $v_k \in \bC$.
We say that $Q$ 
is \emph{radially monotone}
with respect to (w.r.t.) $v_0$
if the distances from $v_0$ to all points of $Q$ are (non-strictly) monotonically increasing.
(Note that requiring the distance to just the vertices of $Q$ to be monotonically increasing
is not equivalent to the same requirement to all points of $Q$.)
We define path $Q$ to be \emph{radially monotone} (without qualification)
if it is radially monotone w.r.t.\ each of its vertices: $v_0,v_1, \ldots, v_{k-1}$.
It is an easy consequence of these definitions that, if $Q$ is radially monotone,
it is radially monotone w.r.t.\ any point $p$ on $Q$, not only w.r.t.\ its vertices.

Before proceeding, we discuss its intuitive motivation.
If a path $Q$ is radially monotone, then ``opening'' the path
with sufficiently small curvatures $\o_i$ at each $v_i$ will avoid
overlap between the two halves of the cut path.
Whereas if a path is not radially monotone, then there is some opening 
curvature assignments $\o_i$ to the $v_i$ that would cause overlap:
assign a small positive curvature $\o_j>0$ to the first vertex $v_j$ at which
radial monotonicity is violated, and assign the other vertices zero or negligible curvatures.
Thus radially monotone cut paths are locally (infinitesimally) opening ``safe,''
and non- radially monotone paths are potentially overlapping.\footnote{
The phrase ``radial monotonicity'' has also appeared in the literature
meaning radially monotone w.r.t.\ just $v_0$, most recently in~\cite{g-aucp-14}. The
version here is more stringent to guarantee non-overlap.}
To further aid intuition, two additional equivalent definitions of radial monotonicity 
are provided in Appendix~\secref{RadialMono.Appendix}

%

\subsection{Angle-monotone chains are radially monotone}
\seclab{angmono=>rm}
Recall the definition of an angle-monotone path from Section~\secref{AngMonoPaths}.
Fig.~\figref{RadialCircles}(c) (in  Appendix~\secref{RadialMono.Appendix}) 
illustrates why a 
$\q$-monotone chain $Q$, for any $\q \le 90^\circ$, is radially monotone:
the vector of each edge of the chain points external to the quarter-circle
passing through each $v_i$. And so the chain intersects the $v_0$-centered
circles at most once (definition~(2)), and the angle $\a(v_i) \ge 90^\circ$ (definition~(3)).
Thus $Q$ is radially monotone w.r.t.\ $v_0$.
But then the argument can be repeated for each $v_i$, for the wedge $W(v_i)$ 
is just a translation of $W(v_0)$.

It should be clear that these angle-monotone chains are special cases of
radially monotone chains. But we rely on the spanning-forest theorem in~\cite{lo-ampnt-17}
to yield angle-monotone chains, and we rely on the unfolding properties of radially monotone
chains from~\cite{o-ucprm-16} to establish non-overlap. 
We summarize in a lemma:
\begin{lemma}
A $\q$-monotone chain $Q$, for any $\q \le 90^\circ$, is radially monotone.
\lemlab{angmono=>rm}
\end{lemma}

\section{Noncrossing $L$ \& $R$ Developments}
\seclab{Noncrossing}
\begin{tcolorbox}
5.~Radial monotonicity of $L$ and $R$, and sufficiently small $\F$,
imply that $L$ and $R$ do not cross in their planar development.
This is a simplified version of a result from~\cite{o-ucprm-16},
and here extended to trees.
\end{tcolorbox}


We will use $\Q = (u_0,u_1, \ldots, u_k)$ as a path of edges
on $\C$, with each $u_i \in \R^3$ a vertex and each $u_i u_{i+1}$ an edge of $\C$.
%
Let $Q=(v_0,v_1,\ldots,v_k)$ be a chain in the plane.
Define the \emph{turn angle} $\t_i$ at $v_i$ to be 
the counterclockwise angle from $v_i-v_{i-1}$ to $v_{i+1}-v_i$.
Thus $\t_i=0$ means that $v_{i-1},v_i,v_{i+1}$ are collinear. 
$\t_i \in (-\pi,\pi)$; simplicity excludes $\t_i = \pm \pi$.

Each turn of the chain $Q$ sweeps out a sector of angles.
We call the union of all these sectors $\L(Q)$; this forms a cone
such that, when apexed at $v_0$, $Q \subseteq \L(Q)$.
The rays bounding $\L(Q)$ are determined by the segments of $Q$ at extreme angles;
call these angles $\s_{\max}$ and $\s_{\min}$.
See Fig.~\figref{RMInduction1} for examples.
Let $|\L(Q)|$ be the measure of the apex angle of the cone, $\s_{\max}-\s_{\min}$.
We will assume that $|\L(Q)| < \pi$ for our chains $Q$, although it is quite possible for
radially monotone chains to have $|\L(Q)| > \pi$. In our case, in fact $|\L(Q)| < \pi/2$, but
that tighter inequality is not needed for Theorem~\thmref{RMInduction} below.
The assumption $|\L(Q)| < \pi$ guarantees that $Q$ fits in a halfplane $H_Q$ whose
bounding line passes through $v_0$.

Because $\s_{\min}$ is turned to $\s_{\max}$, we have that 
the total absolute turn $\sum_i | \t_i | \ge  |\L(Q)|$.
But note that the sum of the turn angles $\sum_i \t_i$ could be smaller than $|\L(Q)|$ because of cancellations.

\subsection{The left and right planar chains $L$ \& $R$}
\seclab{LandR}
Let $\o_i$ be the curvature at vertex $u_i$ of $\Q$.
We view $u_0$ as a leaf of a cut forest, which will then serve as
the end of a cut path, and the ``source'' of opening that path.

Let $\l_i$ be the surface angle at $u_i$ left of $\Q$, and $\r_i$ the surface angle right of $\Q$ there.
So $\l_i + \o_i + \r_i = 2 \pi$, and $\o_i \ge 0$.
Define $L$ to be the planar path from the origin with left angles $\l_i$,
$R$ the path with right angles $\r_i$.
These paths are the left and right planar developments of $\Q$.
(Each of these paths are understood to depend on $\Q$: $L=L(\Q)$ etc.)
We label the vertices of the developed paths $\ell_i,  r_i$.



Define $\o(\Q) = \sum_i \o_i$, the total curvature along the path $\Q$.
We will assume $\o(\Q) < \pi$, a very loose constraint in our
nearly flat circumstances. 
For example, with $\F=30^\circ$, $\O$ for $\C$ is $< \pi \F^2 \approx 49^\circ$,
and $\o(\Q)$ can be at most $\O$.

\subsection{Left-of Definition}
\seclab{LeftOf}
Let $A=(a_0,\ldots,a_k)$ and $B=(b_0,\ldots,b_k)$ be two (planar) radially monotone chains
sharing $x=a_0=b_0$.
(Below, $A$ and $B$ will be the $L$ and $R$ chains.)
Let $D(r)$ be the circle of radius $r$ centered on $x$.
$D(r)$ intersects any radially monotone chain in at most one point
(Appendix~\secref{RadialMono.Appendix}).
Let $a$ and $b$ be two points on $D(r)$.
Say that $a$ is \emph{left of} $b$, $a \preceq b$, if the counterclockwise arc from $b$ to $a$ is less than $\pi$.
If $a=b$, then $a \preceq b$.
Now we extend this relation to entire chains.
Say that chain $A$ is \emph{left of} $B$, $A \preceq B$, if,
for all $r>0$, if $D(r)$ meets both $A$ and $B$, in points $a$ and $b$ respectively,
then $a \preceq b$. If $D(r)$ meets neither chain, or only one, no constraint is specified.
Note that, if $A \preceq B$, $A$ and $B$ can touch but not properly cross.

\subsection{Noncrossing Theorem}
\seclab{NoncrossingThm}

\begin{theorem}
Let $\Q$ be an edge cut-path on $\C$, and $L$ and $R$ the planar
chains derived from $\Q$, as described above.
Under the assumptions:
\begin{enumerate}
\item 
Both $L$ and $R$ are radially monotone,
\item
The total curvature along $\Q$ satisfies $\o(\Q) < \pi$.
\item 
Both cone measures are less than $\pi$:
$|\L(L)| < \pi$ and $|\L(R)| < \pi$,
\end{enumerate}
then $L \preceq R$:
$L$ and $R$ may touch and share an initial chain
from $\ell_0=r_0$, but $L$ and $R$ do not properly cross,
in either direction.
\thmlab{RMInduction}
\end{theorem}
That the angle conditions~(2) and~(3) are necessary
is shown in Appendix~\secref{Noncrossing.Appendix}.

\noindent
\begin{proof}
We first argue that $L$ cannot wrap around as in Fig.~\figref{AngleCondTight}(a)
and cross $R$ from its right side to its left side.
Let $\r_{\max}$ be the counterclockwise bounding ray of $\L(R)$.
In order for $L$ to enter the halfplane $H_R$ containing $\L(R)$,
and intersect $R$ from its right side,
$\r_{\max}$ must turn to be oriented to enter $H_R$, a turn of $\ge \pi$.
We can think of the effect of $\o_i$ as augmenting $R$'s turn angles $\t_i$ to $L$'s turn
angles $\t'_i = \t_i + \o_i$.
Because $\o_i \ge 0$ and $\o(\Q) = \sum_i \o_i < \pi$,
the additional turn of the chain segments of $R$ is $< \pi$,
which is insufficient to rotate $\r_{\max}$ to aim into $H_R$.
See Fig.~\figref{AngleCondTight}(b).
(Later (Section~\secref{Cextension}) we will see that we can assume $L$ and $R$ are
arbitrarily long, so there is no possibility of $L$ wrapping around the end of $R$
and crossing $R$ right-to-left.)

Next we show that $L$ cannot cross $R$ from left to right.
We imagine $Q$ right-developed in the plane, so that $Q=R$.
We then view $L$ as constructed from a fixed $R$ by successively opening/turning the links 
of $R$ by $\o_i$ counterclockwise about $r_i$, with $i$ running backwards
from $r_{n-1}$ to $r_0$, the source vertex of $R$.
Fig.~\figref{RMInduction1}(b) illustrates this process.
Let $L_i= (\ell_i, \ell_{i+1},\ldots,\ell_k)$ be the resulting subchain of $L$ after rotations
$\o_{n-1}, \ldots, \o_i$, and $R_i$ the corresponding subchain of $R = (r_i, r_{i+1},\ldots,r_k)$,
with $\ell_i = r_i$ the common source vertex.
Note that $\ell_i$ and $r_i$ are ultimately not coincident when the chains
are fully opened, but in the proof we can imagine translating $L_i$ so that
$\ell_i = r_i$ without changing the radial monotonicity properties of
either $L_i$ or $R_i$.
We prove $L_i \preceq R_i$ by induction.\footnote{
A ``forward'' proof, from $i=1$ to $i=n-1$, is also possible [Jan.~2021, unpublished].}

$L_{n-1} \preceq R_{n-1}$ is immediate because $\o_{n-1} \le \o(\Q) < \pi$; see Fig.~\figref{RMInduction1}(b).
Assume now $L_{i+1} \preceq R_{i+1}$, and consider $L_i$;
refer to Fig.~\figref{RMInduction2} (Appendix~\secref{Noncrossing.Appendix}).
Because both $L_i$ and $R_i$ are radially monotone, circles centered
on $\ell_i = r_i$ intersect the chains in at most one point each.
Here we are relying on the radial monotonicity properties of $L_i$ and $R_i$
individually, properties they inherit as subchains of the radially monotone $L$ and $R$.
$L_i$ is constructed by rotating $L_{i+1}$ rigidly by $\o_i$ counterclockwise about $\ell_i = r_i$;
see Fig.~\figref{RMInduction2}(b).
This only increases the arc distance between the intersections with those circles,
because the circles must pass through the gap representing $L_{i+1} \preceq R_{i+1}$,
shaded in Fig.~\figref{RMInduction2}(a).
And because we already established that $L$ cannot enter the $R$ halfplane $H_R$, we
know these arcs are $< \pi$: 
for an arc of $\ge \pi$ could turn $\r_{\max}$ to aim into $H_R$.
So $L_i \preceq R_i$.
Repeating this argument back to $i=0$ yields $L \preceq R$, establishing the theorem.
\end{proof}

\begin{figure}[htbp]
\centering
\includegraphics[width=0.7\linewidth]{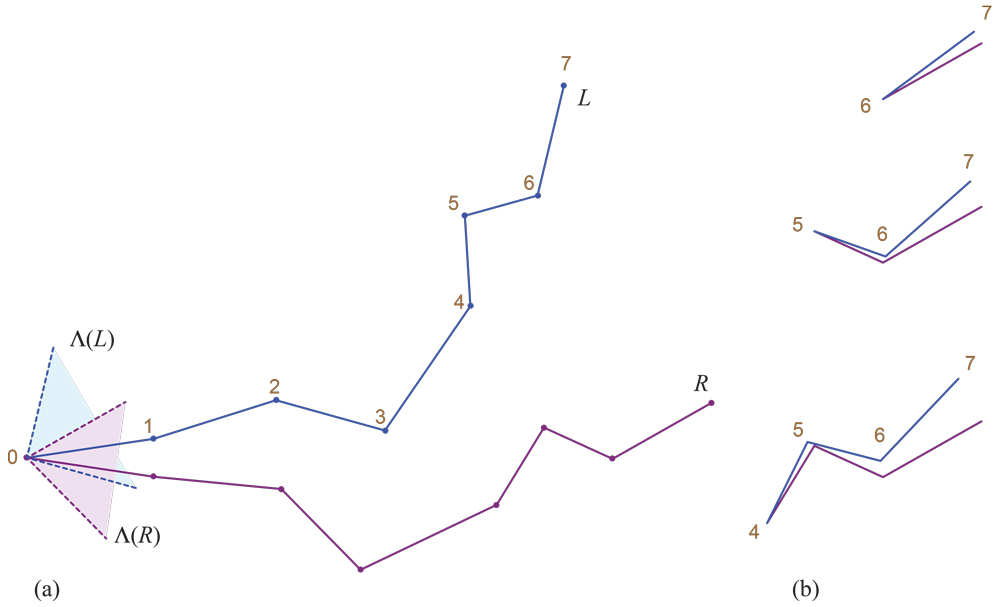}
\caption{(a)
$\o_i = (17^\circ, 6^\circ, 7^\circ, 0^\circ, 5^\circ, 5^\circ, 7^\circ)$, $i=0,\ldots,6$.
(b)~First steps in the induction proof.
See Figure~\protect\figref{RMInduction2} in 
Appendix~\protect\secref{Noncrossing.Appendix} for completion of example.}
\figlab{RMInduction1}
\end{figure}


Our cut paths are (in general) leaf-to-root paths
in some tree $\T \subseteq \cF$ of the forest, so we need to
extend Theorem~\thmref{RMInduction} to trees.\footnote{
This extension was not described explicitly in~\cite{o-ucprm-16}.}
The proof of the following is in Appendix~\secref{Trees.Appendix}.

\begin{corollary}
The $L \preceq R$ conclusion of
Theorem~\thmref{RMInduction}
holds for all the paths in a tree $\T$: $L' \preceq R$,
for any such $L'$.
\corlab{TreePaths}
\end{corollary}


\section{Extending $\C$ to $\C^\infty$}
\seclab{Cextension}
\begin{tcolorbox}
6.~Extending the cap $\C$ to an unbounded polyhedron $\C^\infty$
ensures that the non-crossing of each $L$ and $R$
extends arbitrarily far in the planar development.
\end{tcolorbox}
In order to establish non-overlap of the unfolding, it will help to extend the convex cap $\C$
to an unbounded polyhedron $\C^\infty$ by extending the faces
incident to the boundary $\bcC$. The details are in Appendix~\secref{Cextension.Appendix}.
The consequence is that each cut path $\Q$ can be viewed as extending arbitrarily far 
from its source on $\C$.
This technical trick permits us to ignore ``end effects'' as the cuts are developed in the next section.


\section{Angle-Monotone Strips Partition}
\seclab{Strips}
\begin{tcolorbox}
7.~The development of $\C$ can be partitioned into
$\q$-monotone ``strips,'' whose side-to-side development layout guarantees
non-overlap in the plane.
\end{tcolorbox}
The final step of the proof is to partition the planar $C$ (and so the cap $\C$ by lifting)
into strips that can be developed side-by-side to avoid overlap.
We return to the spanning forest $F$ of $C$ (graph $G$),
as discussed in Section~\secref{SpanningForest}.
Define an \emph{angle-monotone strip} 
(or more specifically, a $\q$-monotone strip) $S$ as a region of $C$ bound
by two angle-monotone paths $L_S$ and $R_S$ which emanate
from the quadrant origin vertex $q \in L_S \cap R_S$, and whose interior is vertex-free.
The strips we use connect from $q$ to each leaf $\ell \in F$, and then follow to the tree's
root on $\bC$.
A simple algorithm to find such strips is described in Appendix~\secref{Waterfall.Appendix};
see Fig.~\figref{Staircasev47_4Q}.
Extending the $\preceq$ relation (Section~\secref{LeftOf})
from curves $L \preceq R$
to adjacent strips, $S_i \preceq S_{i-1}$,
shows that side-by-side layout of these strips
develops all of $\C$ without overlap.
This finally proves Theorem~\thmref{CCapEUnf}.
\begin{figure}[htbp]
\centering
\includegraphics[width=0.5\linewidth]{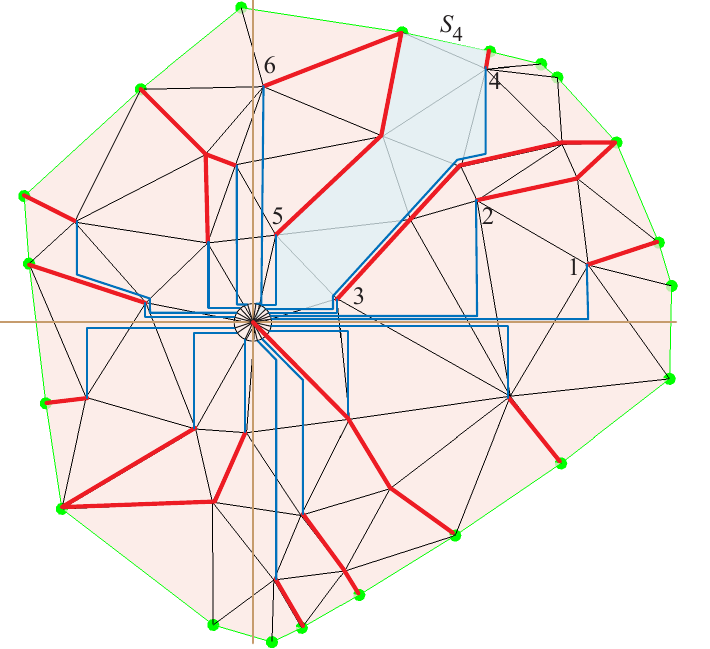}
\caption{Waterfall strips partition. The $S_4$ strip highlighted.}
\figlab{Staircasev47_4Q}
\end{figure}

\section{Discussion}
\seclab{Discussion}
That the polyhedron that results by closing $\C$ with its
convex polygonal base,
can be edge-unfolded without overlap, is proved in~\cite{o-aeucc-17};
see Appendix~\secref{Cap+Base.Appendix}.
I have not pushed on algorithmic time complexity, but certainly $O(n^2)$ suffices;
see Appendix~\secref{Algorithm.Appendix}.

It is natural to hope that Theorem~\thmref{CCapEUnf} can be strengthened.
That the rim of $\C$ lies in a plane is unlikely to be necessary: 
I believe the proof holds as long as shortest paths
from $q$ reach every point of $\bcC$.
Although the proof requires ``sufficiently small $\F$,''
limited empirical exploration
suggests $\F$ need not be that small;
see Fig.~\figref{AM_20_30_s2_v83_ell_Lay3D}.
(The proof assumes the worst-case, with all curvature concentrated on a single path.)
The assumption that $\C$ is acutely triangulated seems overly cautious.
It seems feasible to circumvent the somewhat unnatural projection/lift steps
with direct reasoning on the surface $\C$.

It is natural to wonder\footnote{%
Stefan Langerman, personal communication, August 2017.}
if Theorem~\thmref{CCapEUnf} leads to some type of ``fewest nets'' result
for a convex polyhedron $\P$~\cite[OpenProb.22.2, p.309]{do-gfalop-07}.
At this writing I can only say this is not 
straightforward.
See Appendix~\secref{Discussion.Appendix} for a possible (weak) result.


\subparagraph*{Acknowledgements.}
I benefited from discussions with Anna Lubiw and Mohammad Ghomi.
I am grateful to four anonymous referees, who found an error in Lemma~\lemref{QTurn}
and offered an alternative proof, shortened the justifications
for Lemmas~\lemref{QuadGap} and~\lemref{AngleLifting}, 
suggested extensions and additional relevant references, and improved
the exposition throughout.


\newpage
\bibliography{CCapEUnf}


\newpage
\appendix
\section{Appendix}

\begin{figure}[htbp]
\centering
\includegraphics[width=1.0\linewidth]{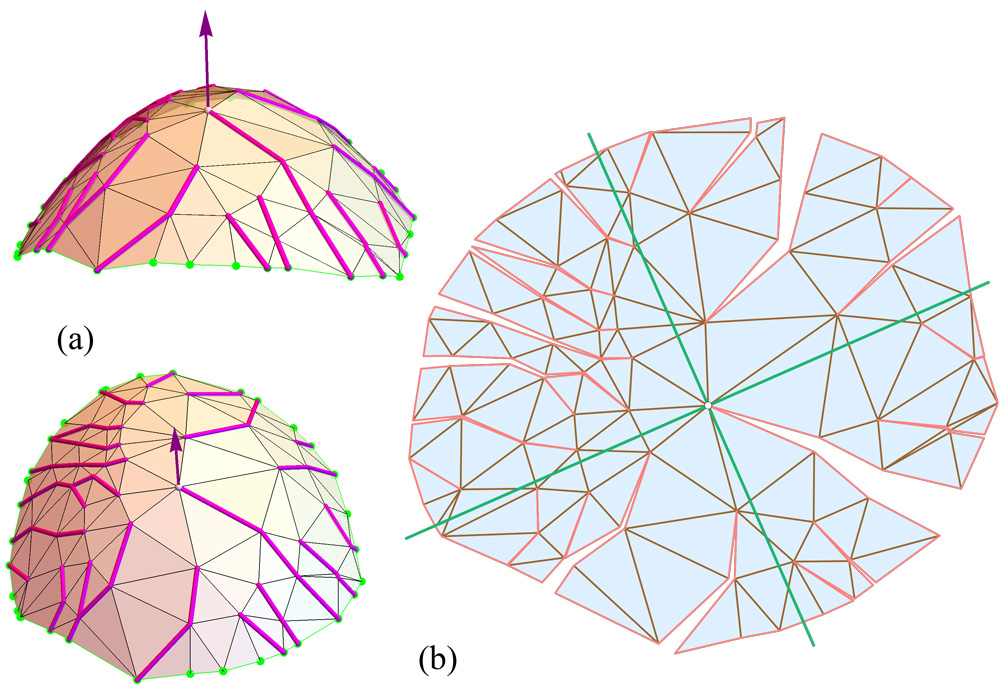}
\caption{(a)~Two views of a convex cap of $83$ vertices with spanning forest $\cF$ marked.
Note $\bcC$ does not lie in a plane. 
$\C$ is non-obtusely triangulated (rather than acutely triangulated).
Here $\F \approx 53^\circ$.
(b)~Edge unfolding by cutting $\cF$.}
\figlab{AM_20_30_s2_v83_ell_Lay3D}
\end{figure}

\subsection{Projection Angle Distortion: Details for Section~\secref{AngDistortion}}
\seclab{AngDistortion.Appendix}
\subsubsection{Notation}
\seclab{NotationDistortion}
Let an angle $\a$ in $\mathbb{R}^3$ be determined by two unit vectors $a$ and $b$.\footnote{
The angle $\a$ in this subsection is unrelated to the acuteness gap introduced
in the Abstract.}
The normal vector $\hat{n} = a \times b$ is tilted $\f$ from the $\hat{z}$-axis,
and spun $\q$ about that axis. 
See Fig.~\figref{AngProj70DeltaMax}. In this figure, $\q$ is chosen to bisect $\a$, which we
will see achieves the maximum distortion.
\begin{figure}[htbp]
\centering
\includegraphics[width=0.5\linewidth]{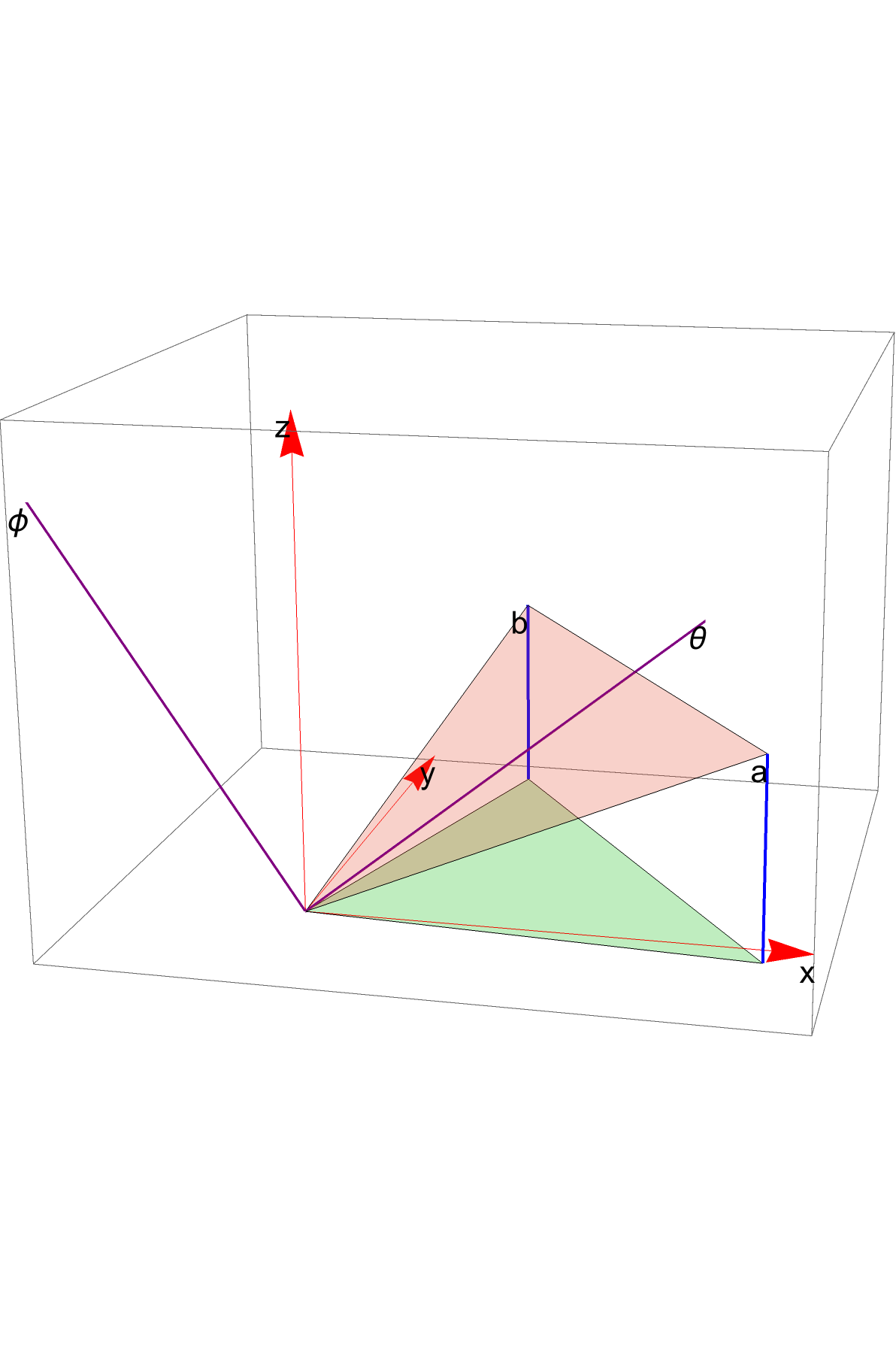}
\caption{$\f=30^\circ$, $\a=70^\circ$ is determined by $a$ and $b$.
The projection to the $xy$-plane (green) results in a larger angle, $\a'=78^\circ$.}
\figlab{AngProj70DeltaMax}
\end{figure}
Let primes indicate projections to the $xy$-plane. 
So $a, b, \a$ project to $a', b', \a'$.
Finally the distortion is $\D_\perp(\a,\f,\q) = |\a'-\a|$; it is about $8^\circ$ in Fig.~\figref{AngProj70DeltaMax}.

\subsubsection{$\q = \a/2$}
\seclab{ThetaAlpha2}
Fixing $\a$ and $\f$, we first argue that the maximum distortion is achieved when
$\q$ bisects $\a$ (as it does in Fig.~\figref{AngProj70DeltaMax}.)

Fig.~\figref{AngProj70Theta} shows $\D_\perp$ as a function of $\q$.
One can see it is a shifted and scaled sine wave with a period of $\pi$.
This remains true over all $\a$ and all $\f$.
\begin{figure}[htbp]
\centering
\includegraphics[width=0.5\linewidth]{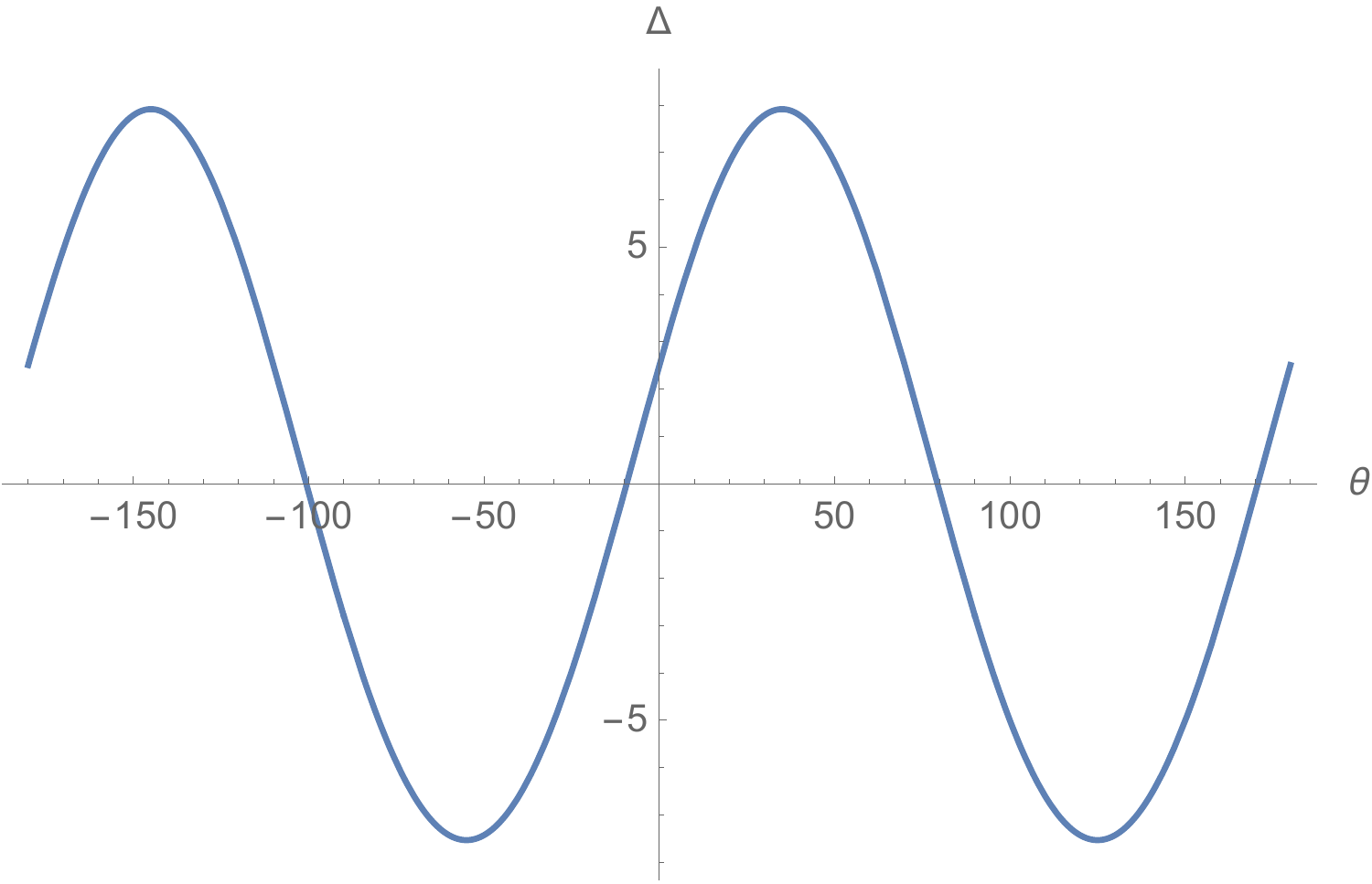}
\caption{For $\f=30^\circ$ and $\a=70^\circ$, the maximum $\D_\perp$ is achieved at $\q=\a/2 = 35^\circ$.}
\figlab{AngProj70Theta}
\end{figure}

\begin{proposition}
For fixed $\f$ and $\a$, the maximum distortion
$\D_\perp(\a,\f,\q)$ is achieved with $\q = \a/2$ (and $\a= \a/2 + \pi$).
\proplab{ThetaAlpha2}
\end{proposition}
We leave this as a claim. 
An interesting aside is that for $\q=\a/2 \pm \pi/4$, $\a'=\a$ and $\D_\perp=0$.

So now we have reduced $\D_\perp(\a,\f,\q)$ to 
depending only on two variables: $\D_\perp(\a,\f)=\D_\perp(\a,\f,\a/2)$.

\subsubsection{Right Angles Worst}
\seclab{RightAngles}
Next we show that the maximum distortion
$\D_\perp(\a,\f)$ occurs when $\a=90^\circ$.
Fig.~\figref{AngProj3D} plots $\D_\perp$ over the full range of $\a$ for a portion of the 
$\f \in [0, 90^\circ]$ range of $\f$.
\begin{figure}[htbp]
\centering
\includegraphics[width=0.75\linewidth]{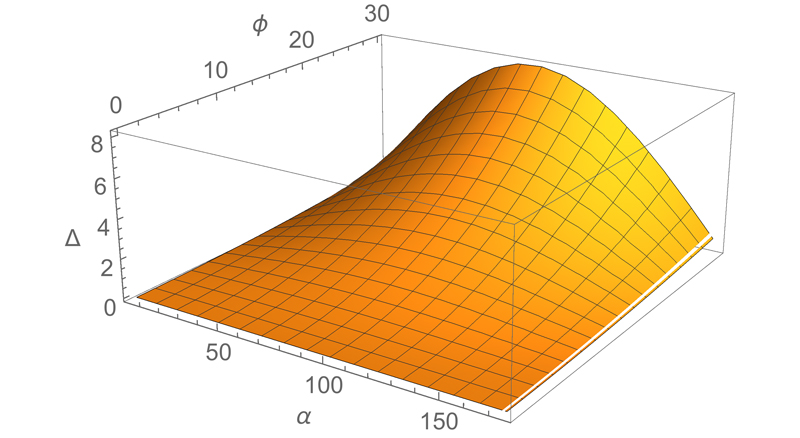}
\caption{$\D_\perp$ as a function of $\a$ and $\f \in [0, 30^\circ]$,
showing the maximum distortion occurs at $\a=90^\circ$.}
\figlab{AngProj3D}
\end{figure}

\begin{proposition}
The maximum distortion $\D_\perp(\a,\f)$ occurs when $\a=90^\circ$,
for all $\f \in [0, 90^\circ]$.
\proplab{RightAngles}
\end{proposition}

\subsubsection{$\D_\perp$ as a function of $\f$}
\seclab{FncPhi}

Propositions~\propref{ThetaAlpha2} and~\propref{RightAngles}
reduce $\D_\perp$ to a function of just $\f$, the tilt of the angle $\a$ in $\mathbb{R}^3$.
Those lemmas permit an explicit derivation\footnote{
This derivation is not difficult but is likely of little interest, so it is not included.} 
of this function:
\begin{equation}
\D_\perp(\f) = \cos^{-1} \left( \frac{ \sin^2 \f }{ \sin^2 \f - 2 } \right) - \pi/2 \;. 
\label{eq:DeltaExact}
\end{equation}
See Fig.~\figref{AngProj90Delta}.
Note that $\D_\perp(0) = \cos^{-1}(0) - \pi/2 = 0$, as claimed earlier.
Thus we have established Lemma~\lemref{Distortion} quoted at the beginning of 
Section~\secref{AngDistortion}.

\begin{figure}[htbp]
\centering
\includegraphics[width=0.75\linewidth]{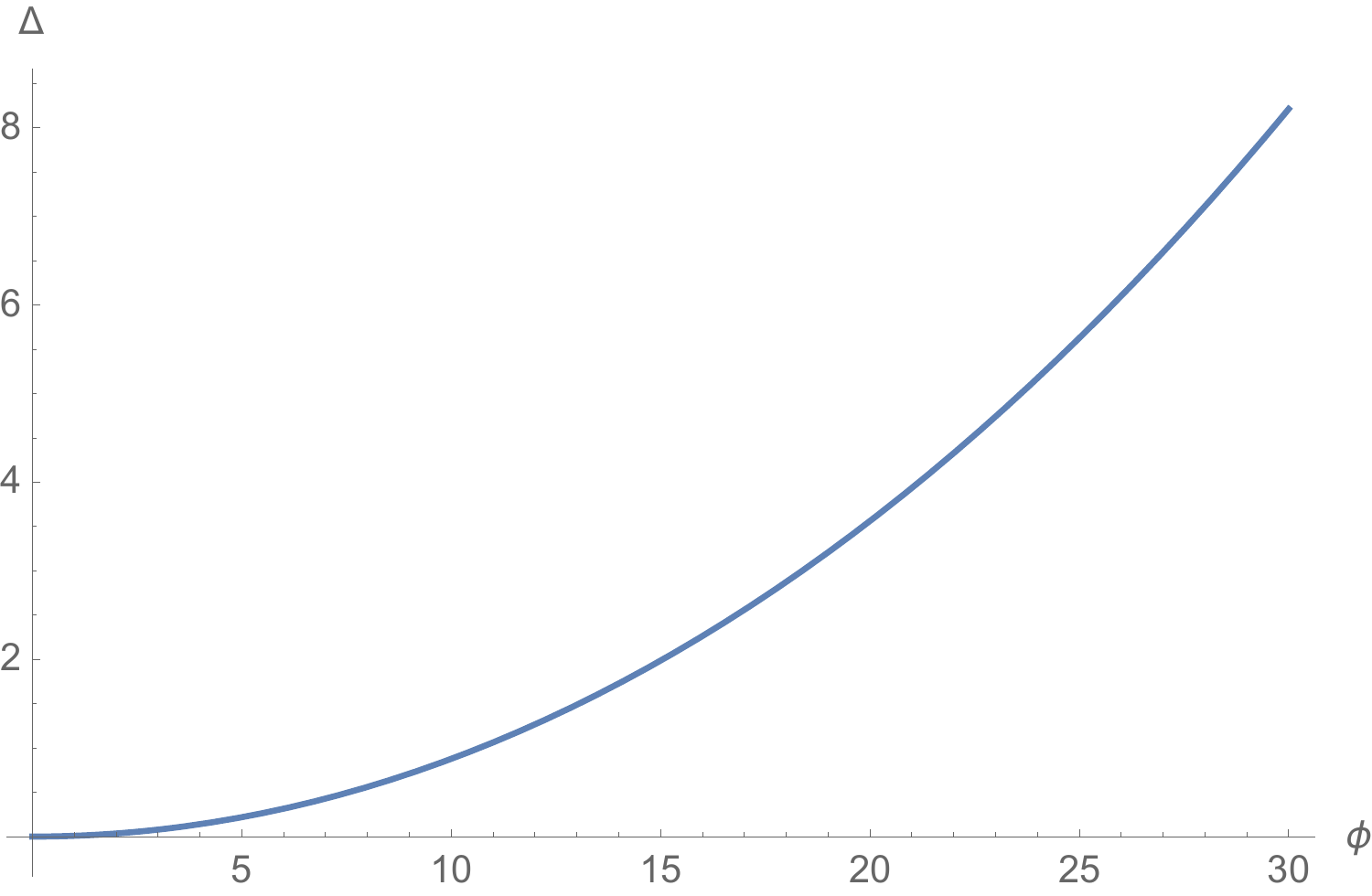}
\caption{$\D_\perp(\f)$.}
\figlab{AngProj90Delta}
\end{figure}

\noindent
For small $\F$, the expression becomes
\begin{equation}
\D_\perp(\F) \approx \F^2 / 2 - \F^4 /12 +O(\F^5) \;.
\tag{\ref{eq:DeltaApprox}}
\end{equation}
A few explicit values:
\begin{eqnarray*}
\D_\perp( 10^\circ ) & \approx & 0.9^\circ \\
\D_\perp( 20^\circ ) & \approx & 3.6^\circ \\
\D_\perp( 30^\circ ) & \approx & 8.2^\circ 
\end{eqnarray*}

\subsection{Curve Distortion: Details on Section~\secref{CurveDistortion}}
\seclab{CurveDistortion.Appendix}
One way to prove Lemma~\lemref{Omega} is to argue that $\O$ is at most
the curvature at the apex of a cone with lateral normal $\F$.
Here we opt for another approach which makes it clear why we cannot
bound $\F$ in terms of $\O$.

The proof depends on the \emph{Gaussian sphere} representation,
a graph $G_S$ on a unit-radius sphere $S$ with nodes corresponding to
each face normal, and arcs corresponding to the dihedral angle of the
edge shared by adjacent faces.
An example is shown in Fig.~\figref{GaussianSphere_n500}.
\begin{figure}[htbp]
\centering
\includegraphics[width=0.75\linewidth]{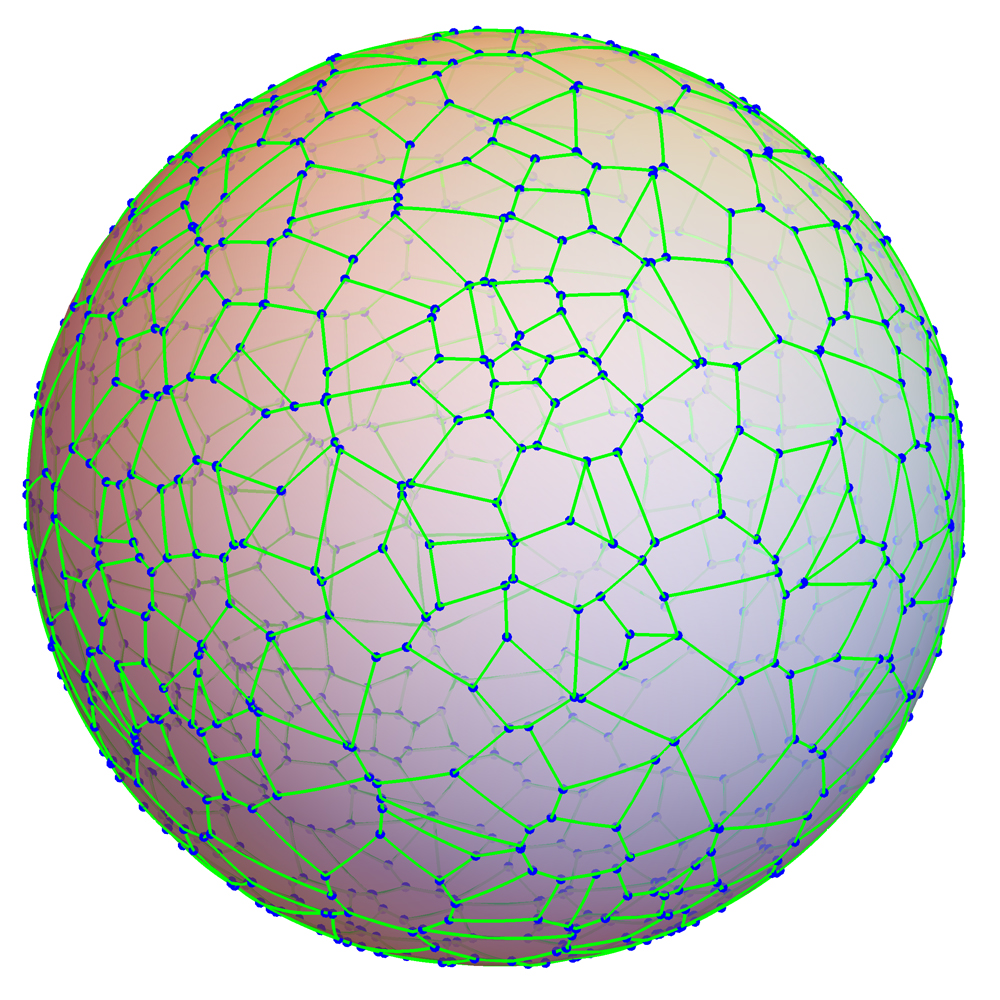}
\caption{Gaussian sphere and $G_S$ for a convex polyhedron of $500$ vertices.}
\figlab{GaussianSphere_n500}
\end{figure}
For a convex polyhedron (and so for a convex cap $\C$), each 
vertex $v$ of $\C$ maps to a convex spherical polygon $s(v)$ whose area is the curvature
at $v$. Each internal angle $\b$ at a face node $f$ of $s(v)$ is $\pi-\a$, where
$\a$ is the face angle incident to $v$ for face $f$.
These basic properties of $G_S$ are well-known; see, e.g.,~\cite{bls-ctelc-07}.

For a vertex $v$, the largest area of its spherical polygon $s(v)$ is achieved when that
polygon approaches a circle. So an upperbound on the area, and so the curvature,
is the area of a disk of radius $\F$. This is the area of a spherical cap,
which is $\O = 2 \pi ( \cos \F -1 )$.
For small $\F$, the area is nearly that of a flat disk, $\pi \F^2$.
This establishes Lemma~\lemref{Omega}.

The reason that we cannot bound $\F$ in terms of $\O$ is that it is possible
that the spherical polygon $s(v)$ is long and thin,
as in Fig.~\figref{Ellipse3DGS_1}. Then its area $\O$ can be small while
the maximum $\F$ deviation is large.
I call such vertices ``oblong''; 
they are revisited in Section~\secref{Discussion.Appendix}.
\begin{figure}[htbp]
\centering
\includegraphics[width=0.75\linewidth]{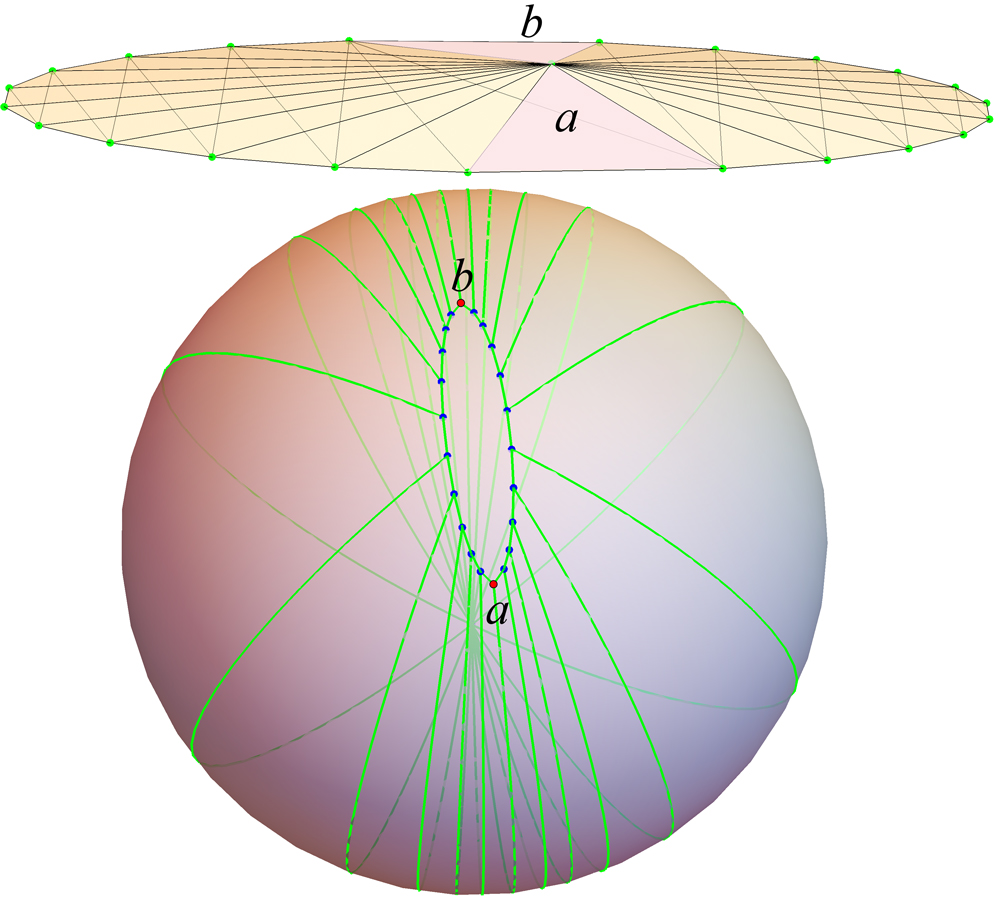}
\caption{Spherical polygon of an ``oblong'' vertex: $| \f(a)-\f(b) |$ is large
but $\O$ is small.}
\figlab{Ellipse3DGS_1}
\end{figure}

For intuition on bounding the curve distortion discussed in 
Section~\secref{CurveDistortion}---and intuition only---we offer
an example in Fig.~\figref{StaircaseDistortion}.
Here an $86^\circ$-monotone staircase $Q'$ lies in the $xy$-plane.
It is lifted to a sphere to $\Q$, with the sphere standing for the convex cap $\C$.
Calculating what could be the angles incident to the left of $\Q$ were
the sphere a triangulated polyhedron, we develop $\Q$ to $\Q_\perp$ in the plane.
$\Q_\perp$ is distorted compared to $Q'$, but by at most $3.8^\circ$, so it remains
$\q$-monotone for $\q < 90^\circ$.
\begin{figure}[htbp]
\centering
\includegraphics[width=0.75\linewidth]{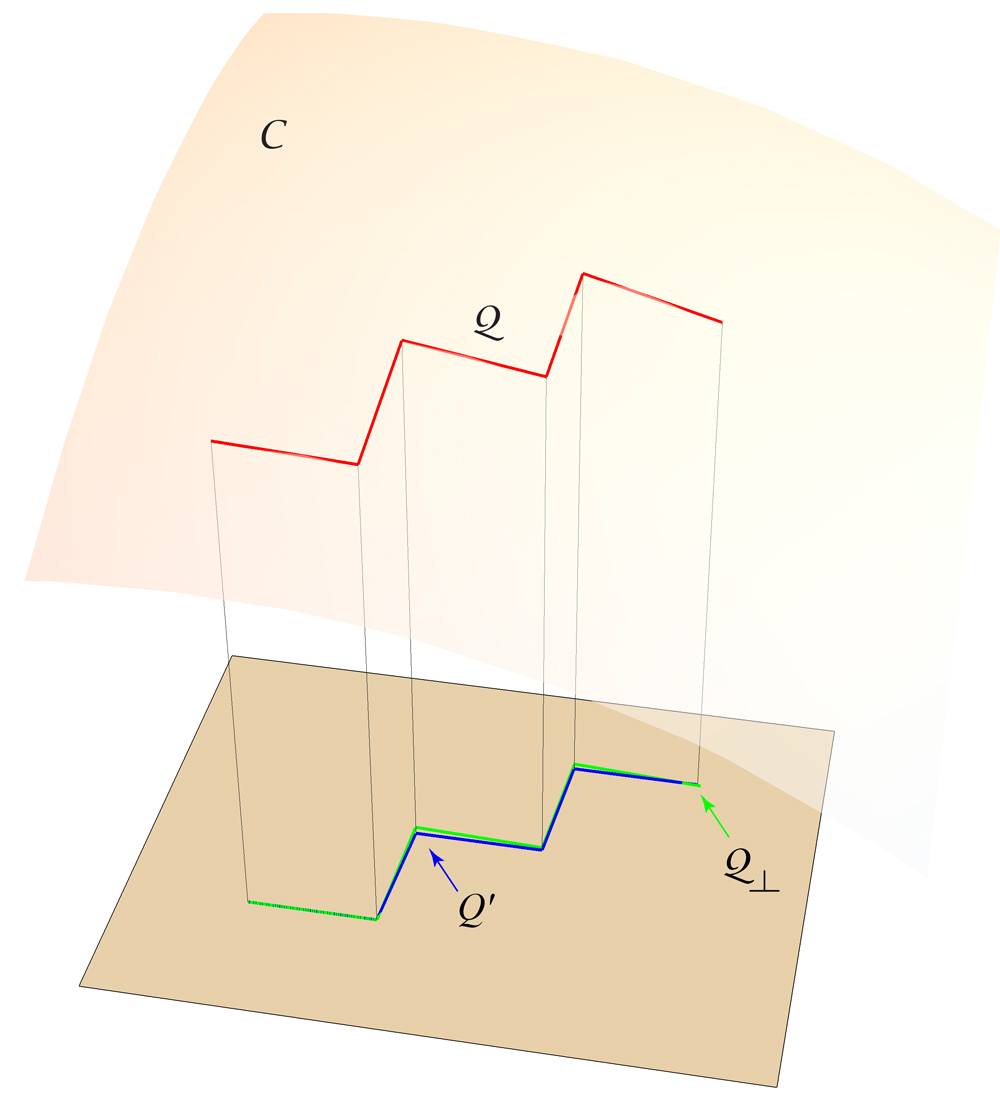}
\caption{$Q'$ is a (blue) path on the plane, $\q$-monotone for $\q=86^\circ$.
Its lift to the sphere is $\Q$ (red).
$\Q_\perp$ (green) is distorted, but remains acute.}
\figlab{StaircaseDistortion}
\end{figure}

\subsubsection{Angle Lifting to $\C$}
\seclab{AngleLifting.Appendix}
We need to supplement the angle distortion calculations presented in Section~\secref{AngDistortion}
for a very specific bound on the total turn of the rims of $C$ and of $\C$.
In particular, here we are concerned with the sign of the distortion.
Repeating from Section~\secref{AngDistortion}, 
let $R' = \bC$ and $R = \bcC$ be the rims of the planar $C$ and of 
the convex cap $\C$, respectively. 

Let $\triangle (a,b,c')$ be a triangle in the $xy$-plane, and $c$ a point vertically above $c'$.
We compare the angle $\psi' = \angle c', a, b$ with angle $\psi = \angle c, a, b$;
see Fig.~\figref{ProjAngs}(a).
(Note that, in contrast to the arbitrary-angle analysis in Section~\secref{AngDistortion},
here the two triangles share a side.)
We start with the fact that the area of the projected $\triangle (c', a, b)$ is 
$\cos \f$ times the area of the 3D $\triangle (c, a, b)$,
where $\f$ is the angle the normal to triangle $\triangle (c, a, b)$ makes with the $\hat{z}$-axis.
Defining $B=b-a$, $C=c-a$ and $C'=c'-a$, we have
\begin{eqnarray*}
B \cdot C' &=& |B| |C'| \cos \psi' \\
B \cdot C &=& |B| |C'| \cos \psi 
\end{eqnarray*}
so
\begin{eqnarray*}
|B| |C'| \cos \psi &=& |B| |C'| \cos \psi'  \cos \f \\
\frac{\cos \psi}{\cos \psi'} &=&  \frac{|C'|}{|C|} \cos \f \\
\mbox{} & \le &  \frac{|C'|}{|C|} \cos \f \\
\mbox{} & \le &  1 \\
\cos \psi & \le & \cos \psi' \\
\psi & \ge & \psi' \;\;\mathrm{when}\;\; \psi' \le 90^\circ \\
\psi & \le & \psi' \;\;\mathrm{when}\;\; \psi' \ge 90^\circ 
\end{eqnarray*}
The last step follows because either both $\psi',\psi \le 90^\circ$ or
$\psi',\psi \ge 90^\circ$.
The conclusion is that the 3D angle $\psi$ is smaller for obtuse $\psi'$,
and larger for acute $\psi'$.
When $\psi'=90^\circ$, then $\psi=90^\circ$.

Now we turn to the general situation of three consecutive vertices $(d,a,b)$ 
of the rim, and how the planar angle at $a$ differs from the 3D angles of
the incident triangles of $\C$.
We start assuming just two triangles are incident to $a$, sharing an edge $ca$.
Note that, if there is no edge of $\C \setminus R$ incident to $a$,
then the triangle $\triangle (d,a,b)$ is a face of $\C$, which implies (by convexity)
that the cap $\C$ is completely flat, and there is nothing to prove.

So we start with the situation depicted in Fig.~\figref{ProjAngs}(b).
We seek to show that the 2D angle $\psi'$ at $a$, $\angle d,a,b$, is always at most
the 3D angle $\psi$, which is the sum of $\psi_1=\angle c,a,b$ and $\psi_2=\angle c,a,d$.
Note that the projection of one of these angles could be smaller and the other larger
than their planar counterparts,
so it is not immediately obvious that the sum is always larger.
But we can see that it as follows.
If $\psi' \le 90^\circ$, then we know that both of the 3D angles $\psi_1$ and $\psi_2$ are
larger by our previous analysis.
If $\psi' \ge 90^\circ$, then partition $\psi' = 90^\circ + \b$, where $\b < 90^\circ$. Then
we have that $\psi_2 = 90^\circ$ and $\psi_1 > \b$, so again $\psi_1 + \psi_2 = \psi \ge \psi'$.

The general situation is that a vertex $a$ on the rim of the cap $\C$
will have several incident edges, rather than just the one $ca$ that we used above.
Continue to use the notation that $d,a,b$ are consecutive vertices of $R'$,
but now edges $c_1, \ldots, c_k$ of $\C$ are incident to $a$.
Consider two consecutive triangles $\triangle (c_{i-1}, a, c_i)$ and $\triangle (c_{i}, a, c_{i+1})$
of $\C$.
These sit over a triangle $\triangle (c_{i-1}, a, c_{i+1})$ which is not a face of $\C$; rather
it is below $\C$ (by convexity).
The argument used above shows that the sum of the two triangle's angles at $a$
are at least the internal triangle's angle at $a$. Repeating this argument shows
that, in the general situation, the sum of all the incident face angles of $\C$ 
is greater than or equal to the 2D angle $\psi' = \angle d,a,b$.

Thus we have proved Lemma~\lemref{AngleLifting} in Section~\secref{CurveDistortion}.


\begin{figure}[htbp]
\centering
\includegraphics[width=0.95\linewidth]{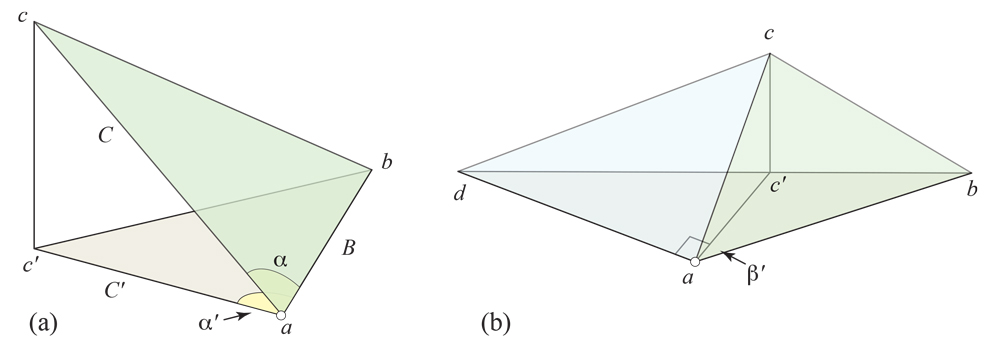}
\caption{(a)~Lifting $\psi'$ to $\psi$ according to a right tetrahedron. 
(b)~Lifting the planar angle at $a$ to the sum of two 3D angles.}
\figlab{ProjAngs}
\end{figure}

\subsubsection{Turn Distortion of $\g'$}
\seclab{TurnDistortion.Appendix}
In order to lead up to the calculation in Section~\secref{TurnDistortion},
we walk through a calculation that will serve as a ``warm-up'' for the calculation
actually needed. I found matters complicated enough to warrant this approach.

Let $\g'$ be 
a simple curve in the $xy$-plane. We aim to bound the total turn difference $\D \g$
between $\g'$ and its lift $\g$ to the cap $\C$.
Let $r' \subset R'$ be the portion of the rim counterclockwise from $b$ to $a$,
so that $\g' \cup r'$ is a closed curve. Of course in the plane there is no curvature enclosed,
and the total turn of this closed curve is $2\pi$.
We describe this total turn $\t'$ in four pieces:
the turn of $\g'$, the turn of $r'$, and the turn at the join points:
\begin{equation}
\t'  =  \t_{\g'} + ( \t_{a'} + \t_{b'} ) + \t_{r'}   \;=\; 3 \pi  \label{eq:turnprime}
\end{equation}
where $\t_{a'}$ and $\t_{b'}$ are the turn angles at $a'$ and $b'$.
See Fig.~\figref{GammaTurn2}(b) (a repeat of Fig.~\figref{GammaTurn}).
\begin{figure}[htbp]
\centering
\includegraphics[width=0.95\linewidth]{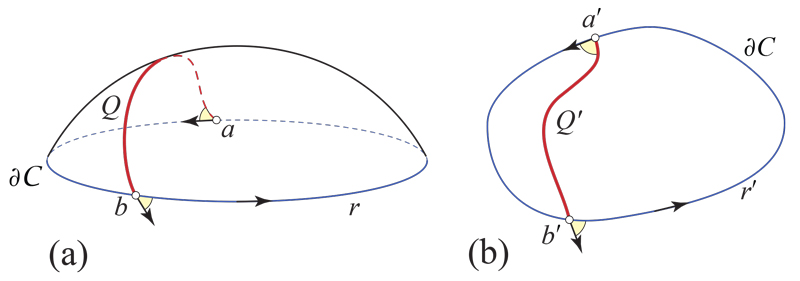}
\caption{(a)~$C$, the projection of the cap $\C$.
(b)~$Q$ is the lift of $Q'$ to $\C$.}
\figlab{GammaTurn2}
\end{figure}

Now we turn to the convex cap $\C$, as illustrated in Fig.~\figref{GammaTurn}(a).
We have a similar expression for $\t$, but now the
Gauss-Bonnet theorem applies: $\t + \o = 2 \pi$, where $\o \le \O$ is the total
curvature inside the path $Q \cup r$:
\begin{equation}
\t  + \o = \t_{Q} + ( \t_{a} + \t_{b} ) + \t_{r} + \o \;=\; 2 \pi \label{eq:turn1}
\end{equation}
Combining Eqs.~\ref{eq:turnprime} and~\ref{eq:turn1},
\begin{eqnarray}
\t_{Q'} + \t_{a'} + \t_{b'} + \t_{r'} & = & \t_{Q} + \t_{a} + \t_{b} + \t_{r} + \o   \\ \nonumber
\t_{Q'}  - \t_{Q} & = &  (\t_{a} - \t_{a'})+ (\t_{b} - \t_{b'}) + (\t_{r}-\t_{r'}) + \o \label{eq:turn2}
\end{eqnarray}
Our goal is to bound $\D Q = |\t_{Q'}  - \t_{Q}|$,
the total distortion of the turn of $Q$ compared to that of $Q'$;
the sign of the distortion is not relevant.

The turn angles at $a$ and $b$ are both distorted by at most $\D_\perp$:
\begin{eqnarray*}
| \t_{a} - \t_{a'} | & \le & \D_\perp \\
| \t_{b} - \t_{b'} | & \le & \D_\perp
\end{eqnarray*}
Note that the analysis in Section~\secref{AngDistortion} shows that the sign of
these angle changes could be positive or negative, depending on whether
$Q'$ meets $R'$ in an acute or obtuse angle. So we bound the absolute magnitude.

From Eq.~\ref{eq:rims}, we have $| \t_{r} - \t_{r'} |  \le   \O$.
Here we do know the sign of the difference, but we only use that sign to 
bound $r \subseteq R$.

Using these bounds in Eq.~\ref{eq:turn2} leads to
\begin{eqnarray*}
\D Q & = & | \t_{Q}  - \t_{Q'} | \\ \nonumber
& \le & 2 \O + 2 \D_\perp  \label{eq:turn3}
\end{eqnarray*}

\paragraph{Example.}
Before moving to the next calculation, we illustrate the preceding with a geometrically accurate
example, the top of a regular icosahedron,
shown in Fig.~\figref{IcosaTop}.
\begin{figure}[htbp]
\centering
\includegraphics[width=0.75\linewidth]{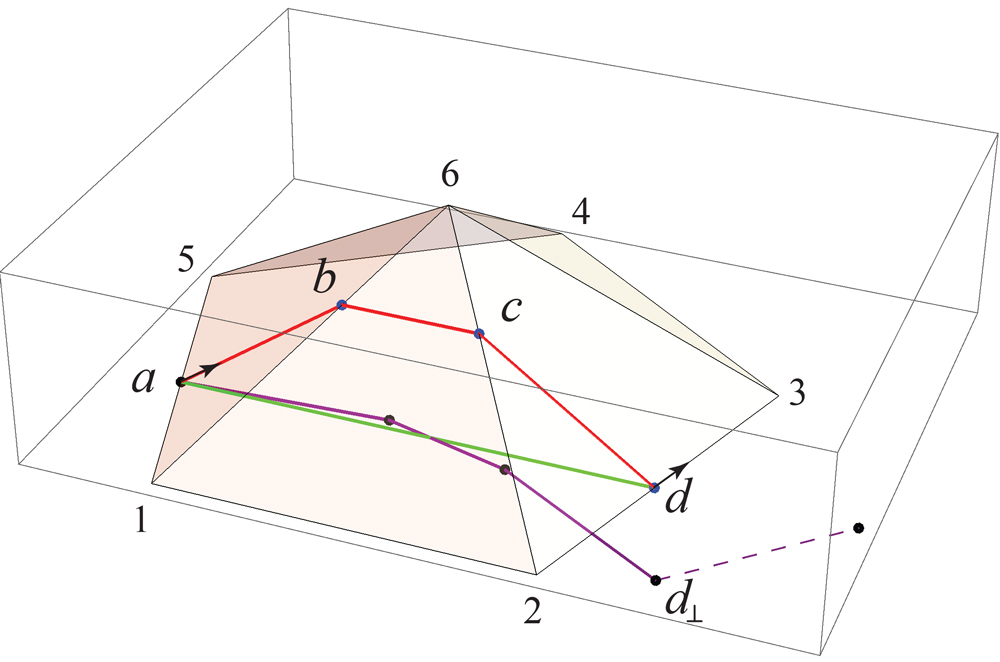}
\caption{Icosahedron cap.
$\Q'=(a,d)$.
$\Q=(a,b,c,d)$.}
\figlab{IcosaTop}
\end{figure}
Here $\F=37.4^\circ$ and $\O=60^\circ$. 
(Lemma~\lemref{Omega} for this $\F$ yields $\O=73.9^\circ$, an upperbound
on the true $\O$.)
$Q'$ is the $(a,d)$ chord of the
pentagon rim, which lifts to $Q=(a,b,c,d)$ on $\C$.
Both the turn at each rim $r'$ vertex, and $\t_{a'} = \t_{d'}$, is $72^\circ$.
So the Gauss-Bonnet theorem for the planar circuit is
\begin{eqnarray*}
 \t_{Q'} + (\t_{a'} + \t_{d'}) + \t_{r'}  & =  & 2 \pi  \\
 0 + (72^\circ + 72^\circ)  + 3 (72^\circ) & =  & 360^\circ
\end{eqnarray*}
The 3D turns $\t_{a} = \t_{d}$ are slightly larger, $75.5^\circ$
(consistent with the analysis in Section~\secref{AngleLifting.Appendix}),
and the turn at each rim $r$ vertex is smaller, $60^\circ$
(consistent with the analysis in Section~\secref{TurnDistortion.Appendix}).
We use the Gauss-Bonnet theorem to solve for $\t_{Q}$:
\begin{eqnarray*}
 \t_{Q} + (\t_{a} + \t_{d}) + \t_{r}  + \o & =  & 2 \pi  \\
 \t_{Q} + (75.5^\circ + 75.5^\circ) + 3 (60^\circ) + 60^\circ & =  & 360^\circ \\
 \t_{Q} & =  & -31.0^\circ 
\end{eqnarray*}
And indeed, $\t_{b}=\t_{c}=-15.5^\circ$.
One can see that the final link of the developed chain $Q_\perp$ has turned $31^\circ$ with
respect to the planar chord,
much smaller than the crude bound of $2 \O + 2 \D_\perp$ derived in the previous section.

\subsection{Radially Monotone Paths: Details for Section~\secref{RadialMono}}
\seclab{RadialMono.Appendix}

Here are two more equivalent definitions of radial monotonicity,
supplementing the one discussed in Section~\secref{RadialMono}.%
\footnote{
In this section we dispense with the primes on symbols to indicate
objects in the $xy$-plane, when there is little chance of ambiguity.
We use $v_i$ for vertices in the plane and $u_i$ for their counterparts
on the cap $\C$.}

\paragraph{(2).}
The condition for $Q$ to be radially monotone w.r.t.\ $v_0$ can be
interpreted as requiring $Q$ to cross every circle centered on $v_0$ at
most once; see Fig.~\figref{RadialCircles}.
The concentric circles viewpoint makes it evident
that infinitesimal rigid rotation of $Q$ about $v_0$ to $Q'$ ensures that
$Q \cap Q' = \{ v_0 \}$, for each point of $Q$ simply moves along its circle.
Of course the concentric circles must be repeated, centered on every vertex $v_i$.
\begin{figure}[htbp]
\centering
\includegraphics[width=0.95\linewidth]{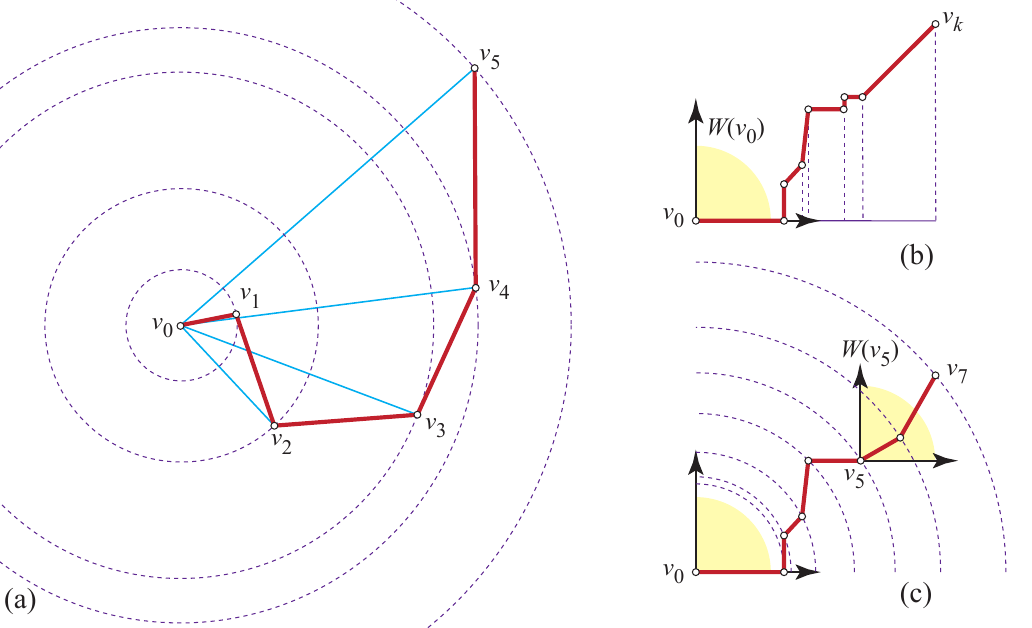}
\caption{(a)~A radially monotone chain, with its
monotonicity w.r.t.\ $v_0$ illustrated.
(b)~A $90^\circ$-monotone chain, with $x$-monotonicity indicated.
(c)~Such a chain is also radially monotone.
}
\figlab{RadialCircles}
\end{figure}

\paragraph{(3).}
A third definition
of radial monotonicity is as follows.
Let $\a(v_i) = \angle (v_0, v_i, v_{i+1})$.
Then $Q$ is radially monotone w.r.t.\ $v_0$ if
$\a(v_i) \ge \pi/2$ for all $i>0$.
For if $\a(v_i) < \pi/2$, $Q$ violates monotonicity at $v_i$,
and if $\a(v_i) \ge \pi/2$, then points along the segment $(v_i, v_{i+1})$
increase in distance from $v_0$.
Again this needs to hold for every vertex as the angle source, not just $v_0$.

Radially monotone paths are the
same\footnote{Anna Lubiw, personal communication, July 2016.}  
as backwards ``self-approaching curves,''
introduced in~\cite{ikl-sac-99} and used for rather different reasons.

Many properties of radially monotone paths are derived in~\cite{o-ucprm-16},
but here we need only the above definitions.

\subsection{Noncrossing $L$ \& $R$ Developments: Details for Section~\secref{Noncrossing}}
\seclab{Noncrossing.Appendix}
Here we show that the angle conditions~(2) and~(3) 
of Theorem~\thmref{RMInduction} are necessary.
Fig.~\figref{AngleCondTight}(a) shows an example where they are violated
and $L$ crosses $R$ from the right side of $R$ to $R's$ left side.
In this figure, $|\L(L)| = |\L(R)| = \pi$, because edge
$r_0 r_1$ points vertically upward and $r_2 r_3$ points vertically downward, and
similarly for $L$. Now suppose that $\o_0=\pi+\e$, and all other $\o_i = 0$.
So $L$ is a rigid rotation of $R$ about $\ell_0=r_0$
by $\o_0$, which allows $L$ to cross $R$
as illustrated.
So some version of the angle conditions are necessary: 
we need that $|\L(R)| + \o(\Q) < 2 \pi$ to prevent this type of ``wrap-around'' intersection,
and conditions~(2) and~(3) meet this requirement.
We address the sufficiency of these conditions in the proof below.
\begin{figure}[htbp]
\centering
\includegraphics[width=0.75\linewidth]{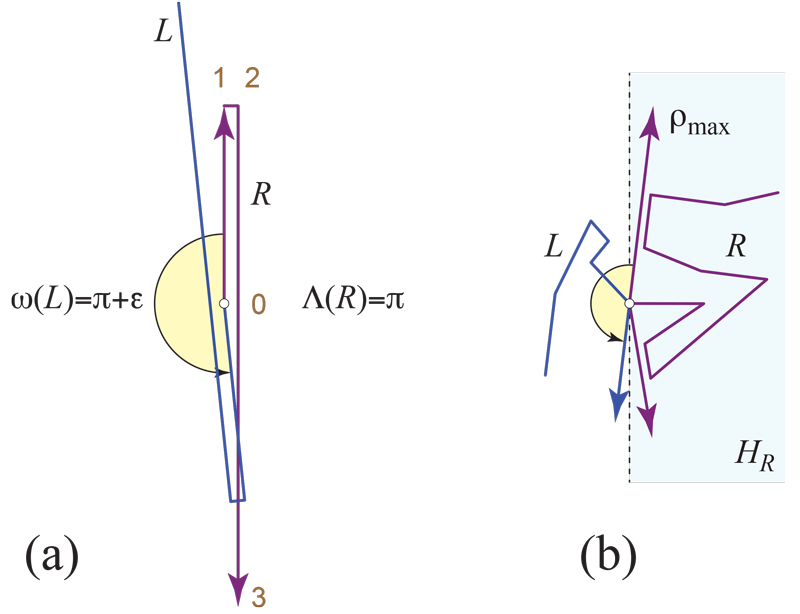}
\caption{(a)~Angle conditions are tight. (b)~Turning $\r_{\max}$.}
\figlab{AngleCondTight}
\end{figure}

The figure below completes Fig.~\figref{RMInduction1}.
\begin{figure}[htbp]
\centering
\includegraphics[width=0.95\linewidth]{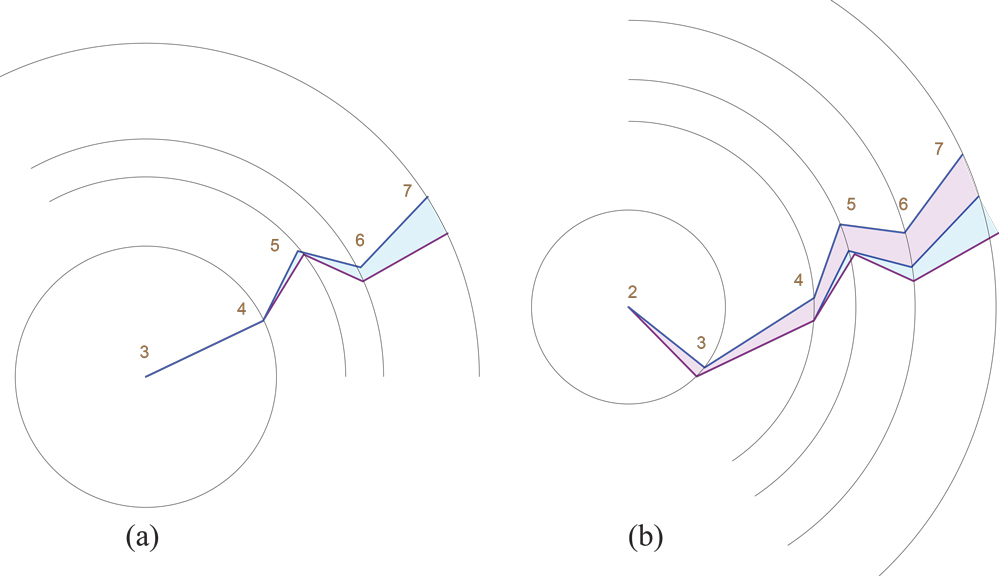}
\caption{Continuation of Fig.~\protect\figref{RMInduction1}.
(a)~$i+1=3$, $L_3 \preceq R_3$. Note: $\o_3 = 0$ so $\ell_3 \ell_4 = r_3 r_4$. (b)~$i=2$, $L_2 \preceq R_2$.}
\figlab{RMInduction2}
\end{figure}
%
A further example is shown in Fig.~\figref{LR_foliation_n12_s1}.
\begin{figure}[htbp]
\centering
\includegraphics[width=0.75\linewidth]{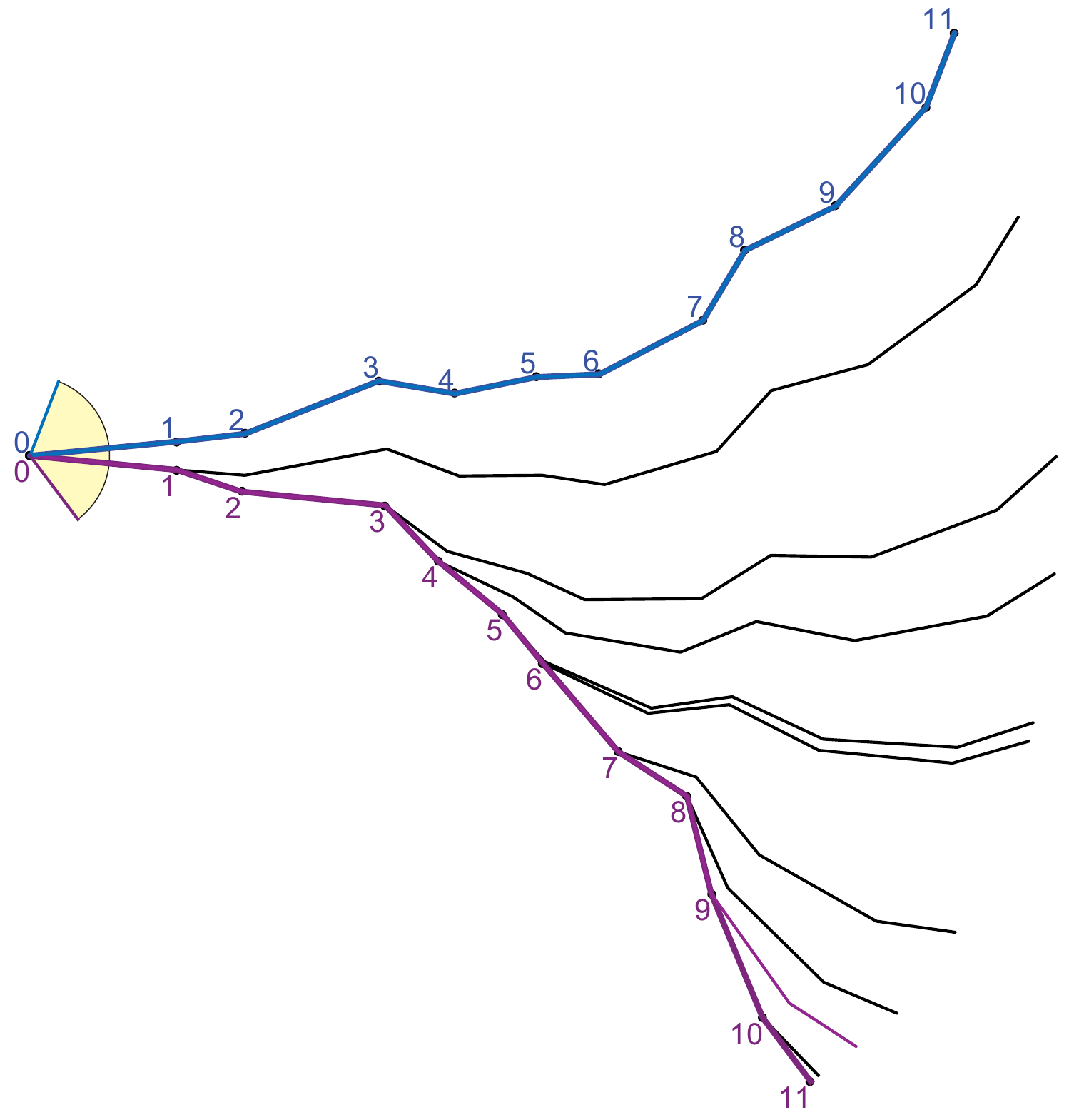}
\caption{$n=11$ example, with $\o( \Q )= 122^\circ$.}
\figlab{LR_foliation_n12_s1}
\end{figure}

\subsubsection{From Paths to Trees: Details for Corollary~\corref{TreePaths}}
\seclab{Trees.Appendix}

Again let $\Q=(u_0,\ldots,u_k)$ be an edge cut-path on $\C$, with $L$ and $R$ the planar
chains derived from $\Q$, just as in Theorem~\thmref{RMInduction}.
Before opening by curvatures, the vertices in the plane are $Q=(v_0,\ldots,v_k)$.
Assume there is path $Q'=(v'_0,\ldots,v'_j=v_i)$ in the tree containing $\Q$,
which is incident to and joins the path at $v_i$ from the left side.
See Fig.~\figref{TreeLeftExpansion}(a).
\begin{figure}[htbp]
\centering
\includegraphics[width=0.75\linewidth]{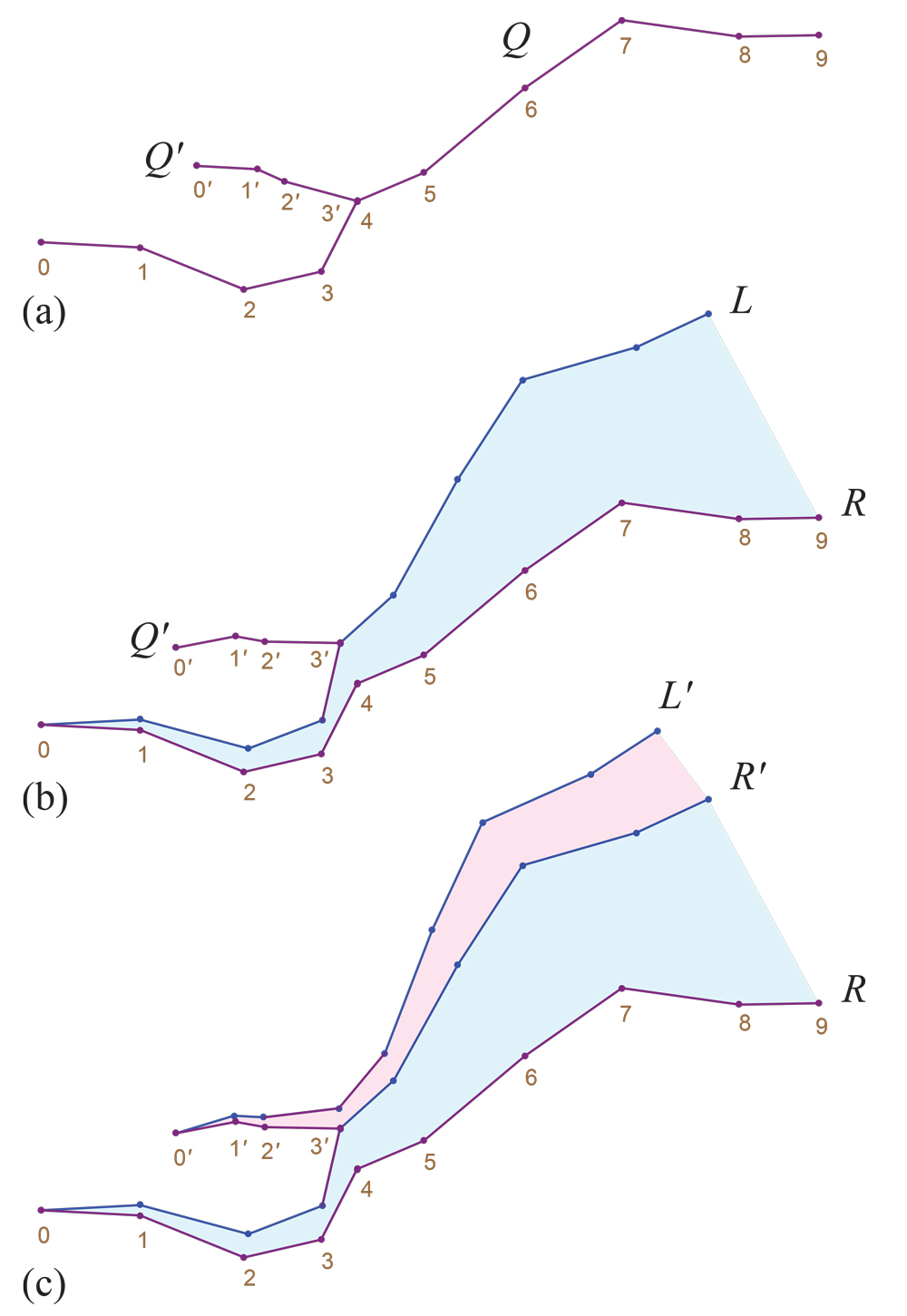}
\caption{(a)~$Q'$ joins $Q$ at $v'_3=v_4$.
(b)~After opening $Q$ to $L$ and $R$.
(c)~After opening $Q'$.}
\figlab{TreeLeftExpansion}
\end{figure}

We fix $Q=R$ and open $Q$ to $L$ as in Theorem~\thmref{RMInduction}, rigidly
moving the unopened $Q'$ attached to $L$ at the same angle $\angle v_{i-1} v_i v'_{k-1}$ at the join;
see Fig.~\figref{TreeLeftExpansion}(b).
Now we apply the same procedure to $Q'$, but now rigidly moving
the tail of $L$, $L_i=(\ell_i,\ldots,\ell_k)$. The logic is that we have already opened that
portion of the path, so the curvatures $\o_i,\ldots,\o_k$ have already been
expended.
See Fig.~\figref{TreeLeftExpansion}(c).

We continue this left-expansion process for all the branches of the tree $\T$,
stacking the openings one upon another.
Rather than assume a curvature bound of
$\o(\Q) < \pi$, we assume that bound summed over the whole tree $\T$: $\o(\T) < \pi$.
For it is the total curvature in all descendants of one vertex $u_i$ that rotates
the next edge $u_i u_{i+1}$.
And similarly, we assume $\L(\T) < \pi$, where $\r_{\max}$ and $\r_{\min}$
range over all edges in $\T$.
These reinterpretations of $\o(\T)$ and $\L(\T)$
retain the argument in Theorem~\thmref{RMInduction} that the total turn of
segments is less than $\pi$, and so avoids ``wrap-around'' crossing of $L'$ with $R$
right-to-left, for any such $L'$.

We can order the leaves of a tree, and their paths to the root,
as they occur in an in-order depth-first search (DFS),
so that the entire tree can be processed in this manner
(this ordering will be used again in Section~\secref{Waterfall.Appendix} below).

\subsection{Extending $\C$ to $\C^\infty$: Details for Section~\secref{Cextension}}
\seclab{Cextension.Appendix}
As mentioned in Section~\secref{Cextension}, it will help to extend the convex cap $\C$
to an unbounded polyhedron $\C^\infty$ as follows.
Define $\C^\infty$ as the intersection of all the halfspaces determined by the faces of $\C$.
Because we have assumed $\F < 90^\circ$, $\C^\infty$ is unbounded.
It will be convenient here to define a ``clipped,'' bounded version of $\C^\infty$:
let $\C^Z$ be $\C^\infty$ intersected with the halfspace $z \ge Z$.
So $\lim_{Z \to -\infty} = \C^\infty$.

We can imagine constructing $\C^Z$ as follows.
Let $B$ be the set of \emph{boundary faces} of $\C$,
those that share an edge with $\bcC$. 
See Fig.~\figref{CCapNormals}(b).
Extend these faces downward.
They intersect one another, and eventually the ``surviving'' faces 
extend to infinity.
We will view $\C^Z$ as $\C \cup \E^Z$, where $\E^Z$ is the
extension ``skirt'' of faces. 
See Fig.~\figref{CCapCZ}.

Note that $\F$ for $\C^\infty$ is the same $\F$ for the original $\C$.
$\C^\infty$ will allow us to ignore the ends of our cuts, as they can be extended arbitrarily
far.

\begin{figure}[htbp]
\centering
\includegraphics[width=0.95\linewidth]{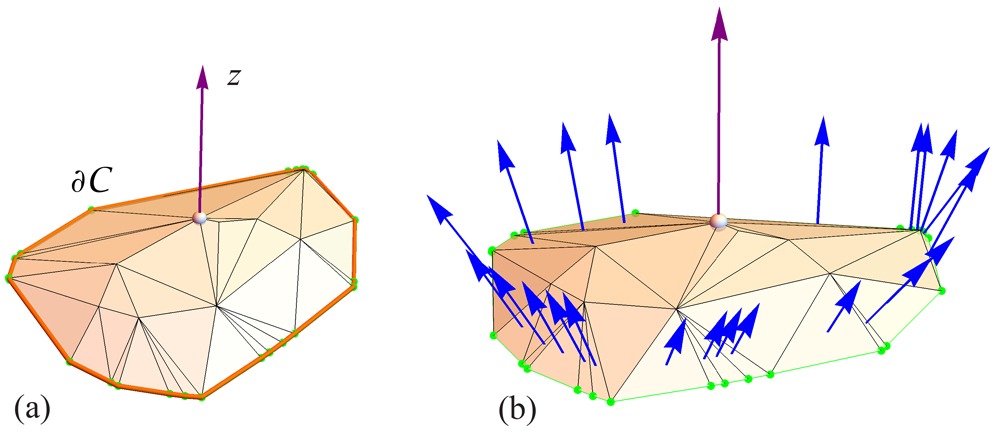}
\caption{(a)~A convex cap $\C$.
(b)~Normals to the boundary faces $B$.
Note: $\C$ is not yet acutely triangulated in this illustration.}
\figlab{CCapNormals}
\end{figure}
\begin{figure}[htbp]
\centering
\includegraphics[width=0.6\linewidth]{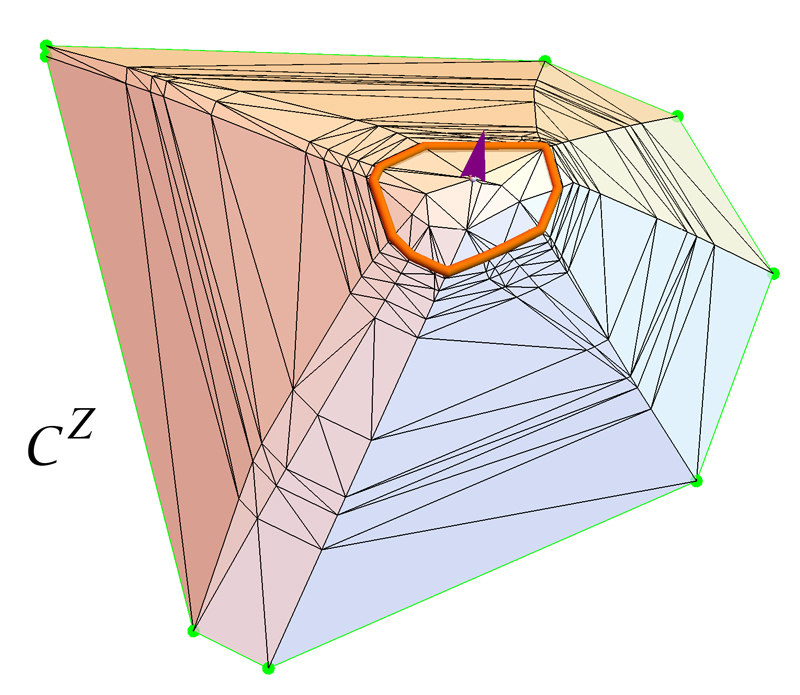}
\caption{Extension of the boundary faces in Fig.~\protect\figref{CCapNormals}(b)
to form $\C^Z$
(not yet acutely triangulated).}
\figlab{CCapCZ}
\end{figure}

%

Recalling that $\C^Z = \C \cup \E^Z$, we need to acutely triangulate $\E^Z$.
We apply Bishop's algorithm, introducing (possibly many) new vertices of curvature zero
on the extension skirt $\E^Z$.
We perform this acute triangulation on the skirt independently of the
triangulation of $\C$, and then glue the two together along $\bcC$.
At the interface $\bcC$, the triangulations may not be compatible,
in that additional vertices may lie along $\bC$, but a path
of $\F$ reaching $\bcC$ can always continue into $\E^Z$
without ever needing to re-enter $C$.
This is illustrated in Fig.~\figref{Incompat_NoReentry}.
\begin{figure}[htbp]
\centering
\includegraphics[width=0.5\linewidth]{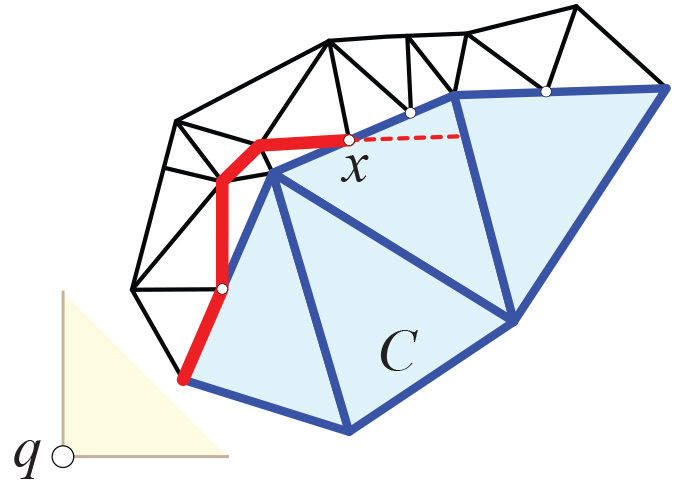}
\caption{The angle-monotone path could follow $\bC$ rather than re-entering $C$ at $x$.}
\figlab{Incompat_NoReentry}
\end{figure}

Note that $\F$ serves as a bound for both $\C$ and $\C^Z$.
The consequence is that each cut path $\Q$ can be viewed as extending arbitrarily far 
from its source on $\C$
before reaching its root on the boundary of $\C^Z$. This permits us to 
``ignore'' end effects as the cuts are developed in Section~\secref{Strips}.

\subsection{Angle-Monotone Strips Partition: Details for Section~\secref{Strips}}
\seclab{Strips.Appendix}
As mentioned, the final step of the proof is to partition the planar $C$ (and so the cap $\C$ by lifting)
into strips that can be developed side-by-side to avoid overlap.\footnote{
We should imagine $\C$ replaced by $\C^Z$, which will make the strips arbitrarily long.}
Define an \emph{angle-monotone strip} 
(or more specifically, a $\q$-monotone strip) $S$ as a region of $C$ bound
by two $\q$-monotone paths $L_S$ and $R_S$ which emanate
from the quadrant origin vertex $q \in L_S \cap R_S$, and whose interior is vertex-free.
The strips we use connect from $q$ to each leaf $\ell \in F$, and then follow to the tree's
root on $\bC$.
For ease of illustration, we will use $\q=90^\circ$, but we will see no substantive modifications
are needed for $\q < 90^\circ$.
Although there are many ways to obtain such a partition, we describe one in particular,
whose validity is easy to see.
We describe the procedure in the $Q_0$ quadrant, with straightforward generalization to the
other quadrants.

\subsubsection{Waterfall Algorithm}
\seclab{Waterfall.Appendix}
Let $T_0$ be the set of leaves of $F$ in $Q_0$, with $|T_0|=n$.
We describe an algorithm to connect each $\ell \in T_0$ to $q$ via noncrossing $\q$-monotone
paths.
Unlike $F$, which is composed of edges of $G$, the paths we describe do not follow edges of $G$.
Consult Fig.~\figref{StaircaseAlgQ0} throughout.
\begin{figure}[htbp]
\centering
\includegraphics[width=0.75\linewidth]{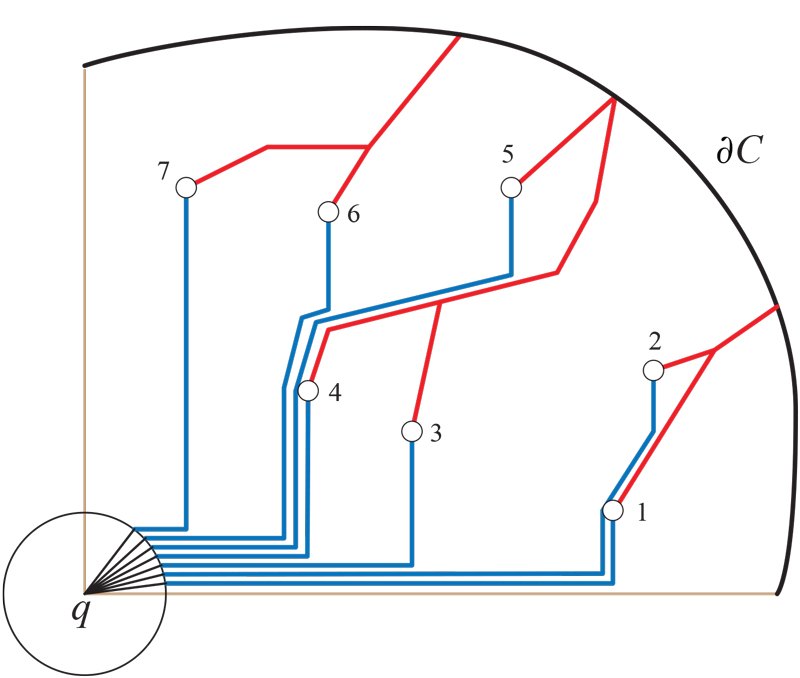}
\caption{Waterfall algorithm in $Q_0$.}
\figlab{StaircaseAlgQ0}
\end{figure}

Center a circle of radius $r$ on the origin $q$, with $r$ smaller than the closest
distance from a leaf to the quadrant axes.
We may assume (Section~\secref{AngMonoForest}) that no vertex aside from $q$ lies
on a quadrant axis, so $r > 0$.
Mark off $n$ ``target points'' $c_i$ on the circle as in Fig.~\figref{StaircaseAlgQ0}.
Process the leaves in $T_0$ in the following order.
Trees in $F$ are processed in counterclockwise order of their root along $\bC$.
Within each tree, the leaves are ordered as they occur in an in-order depth-first search (DFS);
again consult the figure.

Let $y_i$ be the height of $c_i$, $y_1> 0$ and $y_k < r$, and let $\ell_i$ be the $i$-th leaf.
Connect $\ell_1$ to $c_1$ by dropping vertically from $\ell_1$
to height $y_1>0$, and then horizontally to $c_1$. So the connection
has a ``L-shape.'' Then connect radially from $c_1$ to $q$.
Define the path $p_1$ to be this $3$-segment connection joined with the path in $F$ from $\ell_1$ to the root 
on $\bC$.

For the $i$-th step, drop vertically from $\ell_i$ to path $p_{i-1}$,
following just $\e$-above $p_{i-1}$ until it reaches height $y_i$. 
Then connect horizontally to $c_i$ and radially to $q$.
We select $\e$ to ensure noncrossing of the ``stacked'' paths.
It suffices to use $1/(n+1)$ times the minimum of (a)~the smallest vertical distance between a leaf and a point of $F$, and (b)~the smallest horizontal distance between two leaves. 
We also use the same $\e$ to separate $c_{i-1}$ from $c_i$ vertically around the $q$-circle.

To handle $\q < 90^\circ$, all vertical drops instead drop at angle $\q$ inclined with respect to
the $x$-axis.
We take it as clear that each path $p_i$ is $\q$-monotone.
They  are noncrossing because $p_i$ rides above $p_{i-1}$, and $\e$ is small enough
so that $n \e$ cannot bump into a later path $p_j$ for $j>i$.
Thus we have partitioned $C$ into $\q$-monotone strips sharing vertex $q$.
See Fig.~\figref{Staircasev47_4Q} for a complete example.

Returning to Fig.~\figref{StaircaseAlgQ0}, some strips will reach a tree junction before
$\bC$, such as $S_3$ reaching the junction between $\ell_3$ and $\ell_4$.
In that case, the strip continues with the path from that junction to $\bC$, i.e.,
its ``tail'' is a zero-width path.
Other strips, such as $S_4$ in Fig.~\figref{Staircasev47_4Q}, retain non-zero width
to the boundary.

\subsubsection{$L_\perp \preceq R_\perp$}
\seclab{LleftofR}
Define $S'_i$ as the strip counterclockwise of leaf $\ell_i$ in $C'$.
and $L'_{S_i}$ and $R'_{S_i}$ as its right and left boundaries
(which might coincide from some vertex onward).
We reintroduce the primes to distinguish between objects in the planar $C'$,
their lifts on the 3D cap $\C$, and the development from $\C$ back to the plane.
To ease notation, fix $i$, and let $S=S_i$, $L'=L'_{S_i}$, $R'=R'_{S_i}$.
Let $\S$ be the lift of $S'$ to the cap $\C$.
Let $L_\perp$ and $R_\perp$ be the developments of the boundaries of $\S$ back into the plane:
the left development of $R$ and the right development of $L$, so that both developments
are determined by the surface angles within $\S$.
Our goal is to prove $L_\perp \preceq R_\perp$, with the left-of relation $\preceq$ as
defined in Section~\secref{LeftOf}.
We start with proving that $L' \preceq R'$.

By construction, both $L'$ and $R'$ are $\q$-monotone,
and therefore radially monotone. So any circle $D$ centered on $q$ intersects each
in at most one point, say $a'= D \cap L'$ and $b' = D \cap R'$.
The arc from $a'$ to $b'$ lies inside the strip $S$; or perhaps
$a'=b'$ in the tail of $S$.
To prove $L' \preceq R'$,
we only need to show this $a'b'$ arc is at most $\pi$.
This is obvious when $S$ lies in one quadrant.
Although it can be proved if $S$ straddles two quadrants, it is easier to just
use the quadrant boundaries to split a boundary-straddling strip into two halves,
so that always $a'b'$ lies in one quadrant, and so is at most $\pi$.

Now we turn to the lift of $S'$ to $\S$ and the development of the boundaries
$L_\perp$ and $R_\perp$.
Lemma~\lemref{DevelAngMono} guarantees that $L_\perp$ and $R_\perp$
are still $\q$-monotone, for some $\q < 90^\circ$.
Thus the argument is the same as above, establishing that
$L_\perp \preceq R_\perp$ for each strip $\S$.

\subsubsection{Side-by-Side Layout}
\seclab{Side-by-Side}
We now extend the $\preceq$ relation to adjacent strips.
We drop the $\perp$ subscripts, and just let $S_{i-1}$ and $S_i$ be the developments
in the plane of two adjacent strips.
For $S_i \preceq S_{i-1}$ to hold, we require that every circle centered on
their common source $q$ intersects the right boundary of $S_{i-1}$ at point $a$,
the left boundary of $S_i$ at point $b$ counterclockwise of $a$, and intersects no
other strip along the $ab$ arc. 

We just established that the boundaries of both $S_{i-1}$ and $S_i$ are $\q$-monotone,
and so a circle $D$ centered on $q$ will indeed meet the extreme boundaries in
one point each, $a$ and $b$. Here we rely on the extension $\C^\infty$ so that each strip
extends arbitrarily long, effectively to $\infty$.
The two strips share a boundary from $q$ to a leaf vertex $v_i$.
Beyond that they deviate if $\o_i > 0$;
see Fig.~\figref{Strips2}(a).
\begin{figure}[htbp]
\centering
\includegraphics[width=1.0\linewidth]{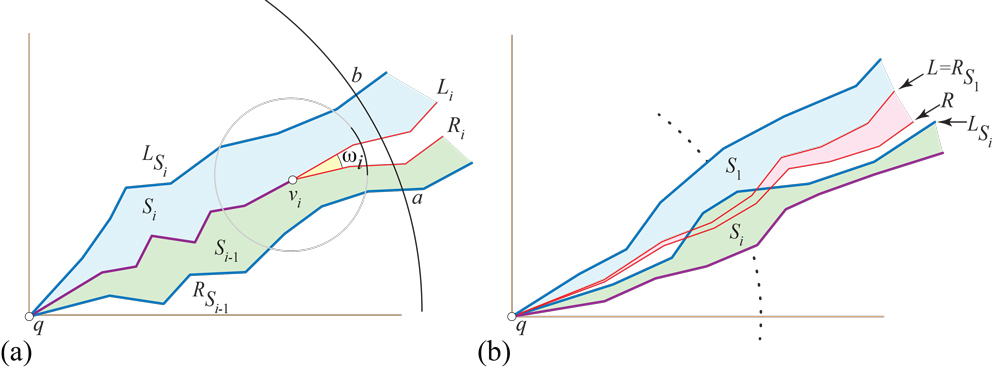}
\caption{(a)~Two adjacent strips. (b)~If $S_i$ crosses into $S_1$,
then $S_n$ crosses into $S_1$, which leads to $L \not\preceq R$.}
\figlab{Strips2}
\end{figure}
We established in Theorem~\thmref{RMInduction} that the two sides of the ``gap''
at $v_i$, $L_i$, and $R_i$, satisfy $L_i \preceq R_i$ (with respect to $v_i$).
This guarantees that there is no surface developed in the gap, which again we
can imagine extending to $\infty$.
So the arc $ab$ crosses $S_{i-1}$, then the gap, then $S_i$.
So indeed $S_i \preceq S_{i-1}$.
Here there is no worry about the length of the arc $ab$ being so long that $L_{S_i}$
could wraparound and cross $R_{S_{i-1}}$
from right-to-left, because each strip fits in a quadrant.

Now we lay out the strips according to their $\preceq$-order.
We choose $S_1$ to be the strip left-adjacent to the gap of the cut
of the forest to $q$ (cf.~Fig.~\figref{QuadGap}), and proceed counterclockwise from there.
The layout of $S_i$ will not overlap with $S_{i-1}$ because $S_i \preceq S_{i-1}$.
Now we argue%
\footnote{
Note here we cannot argue that wraparound intersection doesn't occur because of the
limitation on turning $< \pi$, which only excludes $S_{i}$ wrapping around to $S_{i-1}$.}%
that nor can $S_i$ wraparound and overlap $S_1$.

For suppose $S_i$ crosses into $S_1$ from right-to-left, that is, $L_{S_i}$
crosses $R_{S_1}$;
see Fig.~\figref{Strips2}(b).
Then because $S_{i+1} \preceq S_i$, $S_{i+1}$ must also cross into $S_1$.
Continuing, we conclude that $S_n$ crosses into $S_1$.
Now $S_1$  and $S_n$ are separated by the gap opening of the path in $\cF$ reaching $q$.
The two sides of this gap are $R=L_{S_n}$ and $L=R_{S_1}$.
But we know that $L \preceq R$ by Theorem~\thmref{RMInduction}.
So we have reached a contradiction, and $S_i$ cannot cross into $S_1$.

We continuing laying out strips until we layout $S_n$, which is right-adjacent to the $q$-gap. 
The strips together with that gap fill out the $360^\circ$ neighborhood of $q$.
Consequently, we have developed all of $\C$ without overlap,
and Theorem~\thmref{CCapEUnf} is proved.

\subsection{Algorithm Complexity}
\seclab{Algorithm.Appendix}
Assume we are given an acutely triangulated convex cap $\C$ of $n$ vertices,
and so $O(n)$ edges and faces. The main computation step is finding the
angle-monotone quadrant paths forming a spanning forest in the planar projection $C$, described in 
Section~\secref{AngMonoForest}. This is a simple algorithm, blindly growing paths
until they reach $\bC$ or a vertex already in the forest.
This can be implemented to run in $O(n)$ time.
However, finding the quadrant origin $q$
(cf.~Fig.~\figref{QuadGap}) requires finding shortest paths from vertices to $\bC$.
This can easily be implemented to run in $O(n^2)$ time; likely this can be improved.
All the remainder of the algorithm---developing the cuts---can be accomplished
in $O(n)$ time. So the algorithm is at worst $O(n^2)$, a bound that perhaps could
be improved.

My implementation does not follow the proof religiously.
First, I start with a non-obtuse triangulation rather than an acute triangulation. 
And so I do not tie $\F$ to the acuteness gap $\a$, but instead just use a
reasonably small $\F$.
Second, I do not insist that $\bcC$ lie in a plane.
And, third, I just choose a quadrant origin $q$ near the center of $C$.
It is possible my implementation could lead to overlap,
especially for large $\F$, 
although in my limited testing overlap is easily avoided.

\subsection{Additional Discussion beyond Section~\secref{Discussion}}
\seclab{Discussion.Appendix}
I mentioned that the assumption of acute triangulation seems overly cautious.
The results of~\cite{lo-ampnt-17} extend to $\q$-monotone paths
for widths larger than $90^\circ$ (indeed for any $\q \ge 60^\circ$).
But $\q$-monotone paths for $\q > 90^\circ$ need not be
radially monotone, which is required in Theorem~\thmref{RMInduction}.
This suggests the question of whether the angle-monotone spanning forest could
be replaced with a radially monotone spanning forest.
There is empirical evidence that radially monotone spanning forests
lead to edge-unfoldings of spherical polyhedra~\cite{o-ucprm-16}.
There exist planar triangulations with no radially monotone spanning forest 
(Appendix of~\cite{o-ucprm-16}), but it is not clear they can be realized in $\mathbb{R}^3$
to force overlap. 

Finally, I revisit the fewest nets problem.
As mentioned, it is natural to wonder 
if Theorem~\thmref{CCapEUnf} leads to some type of ``fewest nets'' result
for a convex polyhedron $\P$. 

Define a vertex $v$ to be \emph{oblong} if (a)~the largest disk inscribed in 
$v$'s spherical polygon $s(v)$ (on the Gaussian sphere $G_S$) 
has radius $< \F$, and (b)~the aspect ratio of $s(v)$ is
greater than $2$:$1$.
An example was shown in Fig.~\figref{Ellipse3DGS_1}.
Such oblong vertices neither fit inside nor enclose a $\F$-disk.
If an acutely triangulated convex polyhedron $P$
has no (or a constant number of) oblong vertices, then 
I believe there is a partition of
its faces into a constant number of convex caps that edge-unfold
to nets, with the constant depending on $\F$
(and not $n$).
Unfortunately there do exist polyhedra that have
$\Omega(n)$ oblong vertices.

More particularly, I have a proof outline that, if successful, leads to the following (weak) result:
If the maximum angular separation between face normals incident to any vertex
leads to $\f_{\max}$, and if the acuteness gap $\a$ accommodates $\f_{\max}$ 
according to Eq.~\ref{eq:Dq},
then $\P$ may be unfolded to
$\lesssim 1 / \f^2_{\max}$
non-overlapping nets. 
For example, $n=2000$ random points on a sphere leads to $\f_{\max} \approx 7.1^\circ$
and if $\a \ge 6.9^\circ$---i.e., $\q \le 83.1^\circ$---then $64$ non-overlapping nets suffice
to unfold $\P$.
The novelty here is that
this is independent of the number of vertices $n$.
The previous best result is $\lceil \frac{4}{11}F \rceil= \Omega(n)$ nets~\cite{p-ofnp-07},
where $F$ is the number of faces of $\P$,
which in this example leads to $1454$ nets.
However, the assumption that $\f_{\max}$ is compatible with the acuteness gap $\a$ is
essentially assuming that $\P$ has no oblong vertices.


\subsection{Unfolding Cap-plus-Base Polyhedron}
\seclab{Cap+Base.Appendix}
Here I include just one
illustration from~\cite{o-aeucc-17},
Fig.~\figref{CB_10_20_s3_3D_Unf}, which hints at the construction.
\begin{figure}[htbp]
\centering
\includegraphics[width=1.0\linewidth]{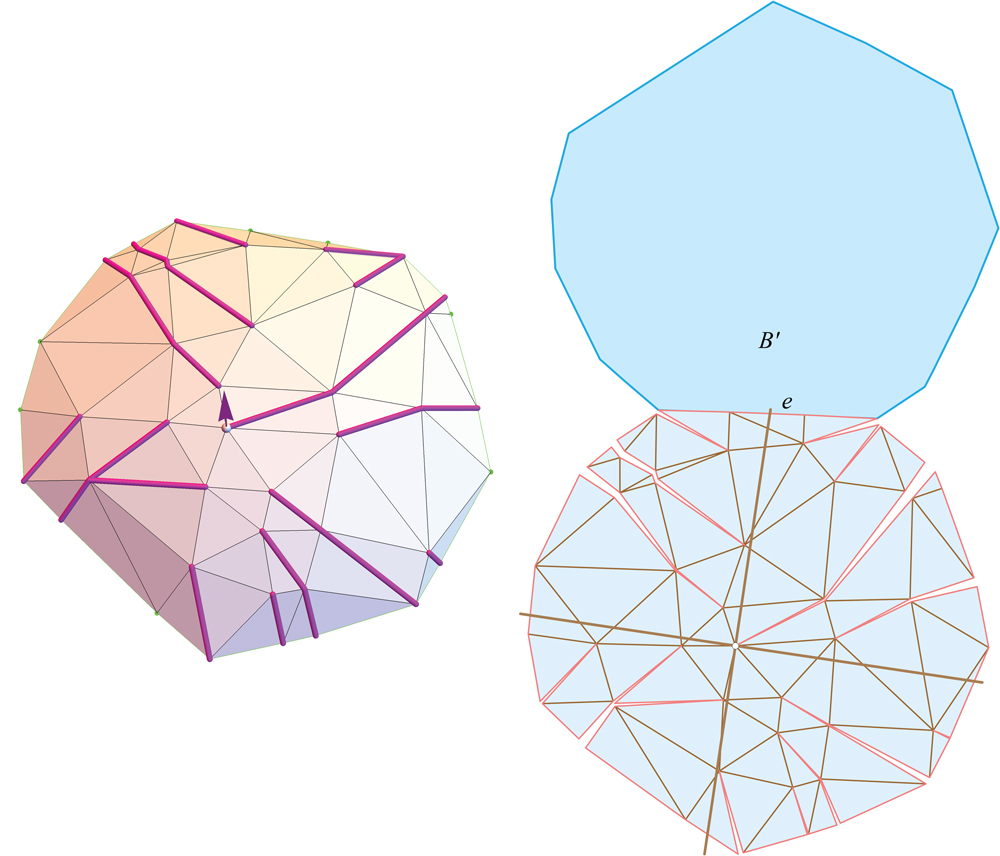}
\caption{Cap $\C$ (left) of $n{=}46$ vertices and an edge-unfolding (right), 
with base $B$ flipped across ``safe edge'' $e$.}
\figlab{CB_10_20_s3_3D_Unf}
\end{figure}

\end{document}